\documentclass[twocolumn,prc,showpacs,preprintnumbers,superscriptaddress]{revtex4-1}

\usepackage[T1]{fontenc}
\usepackage{amsmath} 
\usepackage{amssymb}
\usepackage{amsfonts}
\usepackage{mathrsfs}
\usepackage{units}
\usepackage{multirow}
\usepackage{color}
\usepackage{braket}

\usepackage{dcolumn}
\usepackage{bm}
\usepackage{rotating}

\newcommand{\disregard}[1]{}

\newsavebox{\tmpstrikebox}
\newlength{\tmpstrikelen}

\begin{document}

\title{Gamow-Teller response in the configuration space of DFT-rooted no-core configuration-interaction model}

\author{M. Konieczka}
\affiliation{Institute of Theoretical Physics, Faculty of Physics, University of Warsaw, ul. Pasteura 5,
PL-02-093 Warsaw, Poland}

\author{M. Kortelainen}
\affiliation{Department of Physics, P.O. Box 35 (YFL), University of Jyvaskyla, FI-40014 Jyvaskyla, Finland}
\affiliation{Helsinki Institute of Physics, P.O. Box 64, FI-00014 University of Helsinki, Finland}

\author{W. Satu{\l}a}
\affiliation{Institute of Theoretical Physics, Faculty of Physics, University of Warsaw, ul. Pasteura 5,
PL-02-093 Warsaw, Poland}
\affiliation{Helsinki Institute of Physics, P.O. Box 64, FI-00014 University of Helsinki, Finland}

\date{\today}

\begin{abstract}
\begin{description}
\item[Background]
          The atomic nucleus is a unique laboratory to study fundamental aspects of the electroweak interaction. 
          This includes a question concerning {\it in medium\/} renormalization
          of the axial-vector current, which still lacks satisfactory explanation. Study of spin-isospin or Gamow-Teller (GT) 
          response may provide valuable information on both the quenching of the axial-vector coupling constant 
          as well as on nuclear structure and nuclear astrophysics. 
\item[Purpose]
          We have performed a seminal calculation of the GT response by using the no-core-configuration-interaction
         approach rooted on multi-reference density functional theory (DFT-NCCI). The model treats properly isospin 
         and rotational symmetries and can be applied to calculate both the nuclear spectra and transition rates in atomic nuclei,
         irrespectively of their mass and particle-number parity.           
\item[Methods]
         The DFT-NCCI calculation proceeds as follows:  First, one builds a {\it configuration space\/} 
         by computing relevant, for a given physical problem, (multi)particle-(multi)hole Slater determinants. Next, one applies
         the isospin and angular-momentum projections and performs the isospin- and $K$-mixing in order to construct a 
         {\it model space\/} composed of linearly dependent states of good angular momentum. Eventually, one mixes the projected 
         states by solving the Hill-Wheeler-Griffin equation. 
\item[Results]
         The method is applied to compute the GT strength distribution in selected $N\approx Z$ nuclei including 
         the $p$-shell $^8$Li and  $^8$Be nuclei and the $sd$-shell well-deformed nucleus $^{24}$Mg. 
         In order to demonstrate a flexibility of the approach we present also a calculation of the 
         superallowed GT beta decay in doubly-magic spherical $^{100}$Sn and the low-spin spectrum in $^{100}$In.
\item[Conclusions]
         It is demonstrated that the DFT-NCCI model is capable to capture the GT response satisfactorily well by using relatively
         small {\it configuration space\/} exhausting simultaneously  the GT sum rule.  The model,  due to its flexibility and broad range of applicability,  
         may either serve  as
         a complement or even as an alternative to other theoretical approaches including the conventional nuclear shell model.

\end{description}
\end{abstract}

\maketitle

\section{Introduction}\label{intro}
Single Reference Density Functional Theory (SR-DFT) has proven to be extremely successful in  accounting for the bulk nuclear 
properties like masses, radii, or quadrupole moments over the entire nuclear chart, see \cite{(Ben03),(Erl12)} and 
references quoted therein. The success of SR-DFT or, alternatively, self-consistent mean-field theory has its roots in the spontaneous symmetry 
breaking which allows to incorporate correlations into a single Slater determinant. Deformed wave function 
does not allow, however, for quantum-mechanically-rigorous treatment of neither the nuclear spectra nor the nuclear decay rates. 
So far, this domain was traditionally reserved for the Nuclear Shell Model (NSM), a configuration-interaction (CI) approach involving
strict laboratory-frame treatment of symmetries, see \cite{(Cau05)} for a review. 

An expanse of applicability of the mean-field or Single Reference Energy Density Functional (SR-EDF) based methods 
is ultimately related with symmetry restoration. Recently, strenuous effort was devoted to a development of symmetry-projected multi-reference DFT (MR-DFT) and 
to extend it towards No-Core Configuration-Interaction (NCCI) approach. Bally and coworkers proposed a DFT-NCCI 
framework involving Skyrme superfluid functional and applied it successfully to compute spectra and electromagnetic 
transition rates in $^{25}$Mg~\cite{(Bal14b)}. Our group has developed a variant involving unpaired Skyrme functional and a 
unique combination of angular-momentum and isospin projections and applied to calculate the spectra and beta-decay rates 
in  $N\approx Z$ nuclei from $p$-shell to medium mass nuclei around  $^{62}$Zn \cite{(Sat14),(Sat15a),(Sat16aa),(Kon16)}.  
Recently, the DFT-NCCI method was applied to calculate spectra in neutron-rich $^{44}$S and $^{64}$Cr nuclei with 
Gogny force \cite{(Egi16),(Rod16)}, within relativistic framework \cite{(Zha16)} in $^{54}$Cr, or within pairing-plus-quadrupole model
in magnesium chain \cite{(Bor17)}.

 The results obtained so far have been very promising. In particular, 
they indicate that relatively limited number of configurations is needed to obtain accurate description
of low-energy,  low-spin physics in complex nuclei.  However, further tests of these methods are still required.

The DFT-NCCI method allows to address many important physics questions in a way which is complementary to the conventional NSM. 
The flagship example concerns physical origin of the quenching effect of the weak axial coupling constant (for free-neutron decay $g_{\rm A}=-1.2701(25)$) being a subject of a vivid discussion since the first Gamow-Teller (GT) beta-decay calculations were performed. The DFT-NCCI calculations in $T=1/2$ mirror nuclei \cite{(Kon16)} rather contradict with the statement that {\it the quenching} has its roots in a model space and therefore support the two-body current based explanation, put forward in Refs. ~\cite{(Men11),(Eks14)}, see also \cite{(Bar16),(Leo16),(Klo16)}.

The goal of this work is to compute spin-isospin response by using, for the first time, the DFT-NCCI approach.
The spin-isospin, or GT response,  provides valuable information on  both the electroweak beta decay and nuclear structure. 
Since DFT-NCCI originates from very intuitive and powerful concept of spontaneous symmetry breaking, it gives  a unique opportunity to discuss complex patterns that emerge in the response function in terms of simple deformed single-particle Nilsson levels which are the primary building blocks of 
the formalism. In this sense the DFT-NCCI can be considered again as complementary method with respect to the  NSM~\cite{(Cau05),(Bro01),(Kum16)},
coupled cluster~\cite{(Eks14)}, or
Quasiparticle Random Phase Approximation (QRPA)~\cite{(Nak97),(Sar98),(Eng99),(Mol03), (Paa04a), (Mus16)} which was,  until now,  the only 
possible mean-field-based alternative to the NSM concerning global studies of GT strength distribution. Last but not least, beta decay in $pf$-shell nuclei is studied in variation-after-projection Excited Vampir approach with G-matrix-driven realistic effective interaction ~\cite{(Pet11)}. Although the method is based on a mean-field concept, its model space and treatment of correlations are entirely different 
from the DFT-NCCI model.

This paper is organized as follows. In Sec.~\ref{sec:ncci} we discuss the foundations of the DFT-NCCI model 
paying special attention to the concept of configuration and model spaces. In Sec.~\ref{sec:A8} we present the results for the structure and 
GT strength distribution in $A=8$ nuclei. In Sec.~\ref{sec:sdSD} we discusses  the spin-isospin response in the $sd$-midshell nucleus $^{24}$Mg. Eventually, in Sec.~\ref{sec:A100}, we focus on the $^{100}\textrm{Sn}$ $\to$ $^{100}\textrm{In}$ superallowed GT beta decay and the low-spin spectrum of $^{100}\textrm{In}$. 
Summary and conclusions are presented in Sec.~\ref{sec:conc}.  
All calculations presented in this work were done using developing version of the HFODD solver~\cite{(Sch17)} equipped with the NCCI module.

\section{The DFT-rooted no-core-con\-fi\-gu\-ra\-tion-in\-te\-rac\-tion model}\label{sec:ncci}

The DFT-NCCI models are post Hartree-Fock(-Bogliubov) approaches which mix non-orthogonal many-body states projected from 
symmetry breaking mean-field solutions.  Their sole ingredients are therefore independent-particle (quasi)particle-(quasi)hole configurations 
and projection techniques that are used to restore spontaneously broken symmetries.  In practical applications, the projections are handled by using the generalized Wick's theorem (GWT) which leads from a SR to multi-reference (MR) formulation of the DFT (MR-DFT).  

The GWT allows to handle theory numerically, however it leads to singular kernels once modern density-dependent Skyrme or Gogny 
forces are used for the beyond-mean-field part of the calculation. The intensive work to overcome 
this problem of projection-induced singularities is currently under way. The attempts to regularize the kernels \cite{(Ben09),(Sat14b)} 
have not provided a satisfactory solution so far.  Hence, at present, the theory can be safely carried on only for true interactions like 
the SV$_{\rm T}$ \cite{(Bei75)}, used in the present work, or the SLyM0 \cite{(Sad13)}, which both are density-independent 
Skyrme pseudopotentials.  In addition to these, recently developed regularized finite-range pseudopotential~\cite{(Ben17)}
aims also for beyond-mean-field calculations.
It is worth mentioning that these  pseudopotentials are characterized by anomalously low effective 
mass which affects the single-particle (s.p.) level density and, in turn,  influences spectroscopic properties of the calculated nuclei. 

The MR-DFT approach developed by our group is unique in the sense that it restores angular momentum and treats 
rigorously the isospin symmetry i.e. is retaining only physical sources of its breaking. It provides wave functions which
are isospin ($T$) and $K$ (projection of angular momentum onto intrinsic $z$-axis) mixed as
\begin{equation}
\ket{\varphi; I M; T_z}^{(i)}=\frac{1}{\sqrt{\mathcal{N}^{(i)}_{\varphi;IM;T_z}}}
\sum_{\substack{K, \\ T\geq |T_z|}} a_{KT}^{(i)} \hat{P}^T_{T_zT_z}\hat{P}^{I}_{MK}\ket{\varphi} \,,  \label{eq:mrdftstate}
\end{equation} 
where $\hat{P}^{I}$ and $\hat{P}^{T}$ are projection operators of SU(2) group generated by angular-momentum and isospin, respectively, and $\mathcal{N}^{(i)}_{\varphi;IM;T_z}$ is a normalization constant. 
Index $i$ enumerates different solutions of a given spin $I$. The Slater determinant, $\varphi$, is calculated self-consistently by
using Hartree-Fock  (HF) method with the SV$_{\rm T}$ Skyrme and Coulomb forces. 

The MR-DFT wave functions (\ref{eq:mrdftstate}) can be successfully 
used to compute, for example, beta-decay transition rates between the ground states as shown in Refs.~\cite{(Sat12),(Kon16)}. 
In order to account for beta-decay strength distribution, the MR-DFT concept needs to be extended by including 
the states (\ref{eq:mrdftstate}) projected from many Slater determinants $\varphi_j$ corresponding to
different (multi)particle-(multi)hole excitations. The projected states, which are generally non-orthogonal to each other,  are mixed by solving the 
Hill-Wheeler-Griffin equation with, typically, the same Hamiltonian that was used to generate them at the HF stage \cite{(RS80)}. 
In effect, one obtains a set of linearly independent 
DFT-NCCI eigenstates of the form of
\begin{equation}
\ket{\psi_{\textrm{NCCI}}^{k; IM; T_z}}
=\frac{1}{\sqrt{\mathcal{N}^{(k)}_{IM;T_z}}}
\sum_{ij}c_{ij}^{(k)}\ket{\varphi_j;I M;T_z}^{(i)} \,, \label{eq:nccistate}
\end{equation}  
together with the corresponding energy spectrum. 
More details concerning our method can be found in Ref.~\cite{(Sat16aa)}.

Contrary to the standard NSM, the model space of our DFT-NCCI approach is not fixed.  It is built step by step, by adding physically relevant 
low-lying particle-hole (p-h) mean field configurations which correspond to self-consistent HF solutions conserving parity and signature symmetries. 
The basic idea is to explore all relevant single-particle Nilsson levels.  Hence, in even-even nuclei, we include in the first place the ground-state 
configuration and low-lying aligned ($\ket{h} \otimes \ket{\tilde{p}}$ or $\ket{\tilde{h}} \otimes \ket{p}$) and anti-aligned ($\ket{h} \otimes \ket{p}$ or $\ket{\tilde{h}} \otimes \ket{\tilde{p}}$) 1p-1h configurations where 
$\ket{p}$ and $\ket{\tilde{p}}$ ($\ket{h}$ and $\ket{\tilde{h}}$) label single-particle (single-hole) states of opposite signature. 
In an odd-A nuclei we explore first configurations built by exciting the unpaired nucleon within a fixed signature block. 
In the second step we test stability of the predictions with respect to low-lying broken-pair configurations.   
Similar strategy is used in odd-odd nuclei. In this case, however, one has to consider both aligned and anti-aligned configurations.

\begin{figure}[htb]
\centering
\includegraphics[width=1.0\columnwidth]{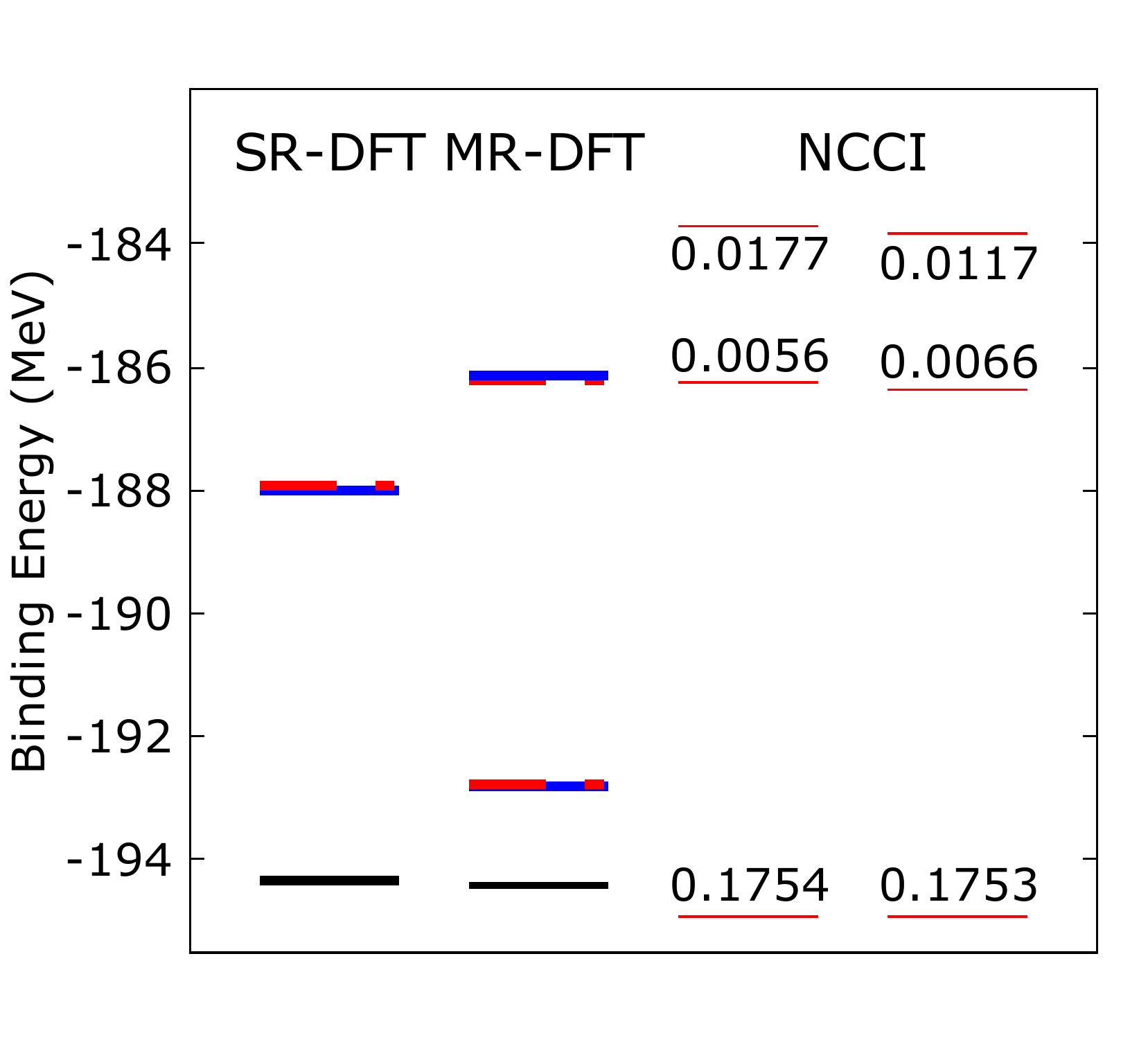}
\caption{(Colour online) Leftmost column shows the HF energies of the g.s. and the lowest $\nu$p-$\nu$h (red, dashed line) and $\pi$p-$\pi$h (blue, solid line)  excitations in $^{24}$Mg.
The second column illustrates $4^+$ states projected from these SR configurations without configuration mixing. 
Last two columns depict the DFT-NCCI results involving different configurations.  Left (right) part shows the CI results involving the g.s. and 
$\nu$p-$\nu$h ($\pi$p-$\pi$h) configurations, respectively.  Numbers in last two columns indicate the calculated GT matrix elements for 
$|\, ^{24}$Al; $4_1^{+} \rangle \rightarrow|^{24}$Mg; $4_i^{+} \rangle$ decay.}
\label{fig:iso}
\end{figure}

In most of the applications, isospin symmetry restoration allows to reduce the configuration space in $N=Z$ nuclei by a 
factor of two due to similarity between the neutron and proton 1p-1h excitations.  The effect is illustrated 
in Fig.~\ref{fig:iso} for a representative example of $^{24}$Mg. In the present calculation, SR ground state (g.s.) 
and the lowest proton ($\pi$p-$\pi$h) and neutron ($\nu$p-$\nu$h) 1p-1h HF configurations were taken into account. 
Energies of excited states differ by 80\,keV  as shown in  
the left column of Fig.~\ref{fig:iso}. By applying the angular-momentum and isospin projections with $I=4^+$, one obtains 
the corresponding, symmetry restored, $I=4_1^+$ from the ground state
and four almost doubly-degenerated excited $I=4^+$ states as shown in the second column of Fig.~\ref{fig:iso}. 
The third and fourth columns show the DFT-NCCI results. The third column depicts configuration mixing calculation
involving two HF configurations, the g.s. and the lowest neutron 1p-1h excitation.  The fourth column shows the results 
of three-configurations mixing, including, in addition to the previous case, the lowest proton 1p-1h excitation. 
Addition of the proton 1p-1h configuration almost does not influence neither 
the spectrum nor the GT matrix elements for  $|\, ^{24}$Al; $4_1^{+} \rangle \rightarrow$ 
$|^{24}$Mg; $4_i^{+} \rangle$ decay.

\section{Low-energy spectra and Gamow-Teller beta decay for A=8 nuclei}\label{sec:A8}

In this section we will investigate the structure and beta-decay properties of very light nuclei 
$^{8}$Be, $^{8}$Li, and $^{8}$He by using the DFT-NCCI framework. The $p$-shell nuclei offer
an excellent playground to test, in particular, a configuration-space dependence of our scheme. 
One should bear in mind, however, that light nuclei are weakly bound. Hence, they may exhibit a 
variety of phenomena which either emerge or strongly depend on the coupling to continuum \cite{(Bla08),(Rii14)}
which is beyond our approach. These effects include clustering, appearance of low-lying broad resonances 
or particle-decay channels that may compete with beta decay and, in turn, significantly influence 
beta-decay strength distribution.

We shall focus on GT strength distributions of $^{8}$He,$0^{+}_{\text{g.s.}}$ and $^{8}$Li,$2^{+}_{\text{g.s.}}$ beta-decays
paying special attention to physical interpretation of particular peaks. For the first time these peaks can be interpreted 
in terms of deformed Nilsson states and deformed Nilsson configurations used in the mixing. We shall also investigate the saturation of GT sum rules
for the lowest $1^{+}$, $2^{+}$, and $3^{+}$ states in $^{8}$Li in order to verify the completeness of the 
model space.

\begin{figure}[htb]
\centering
\includegraphics[width=1.0\columnwidth]{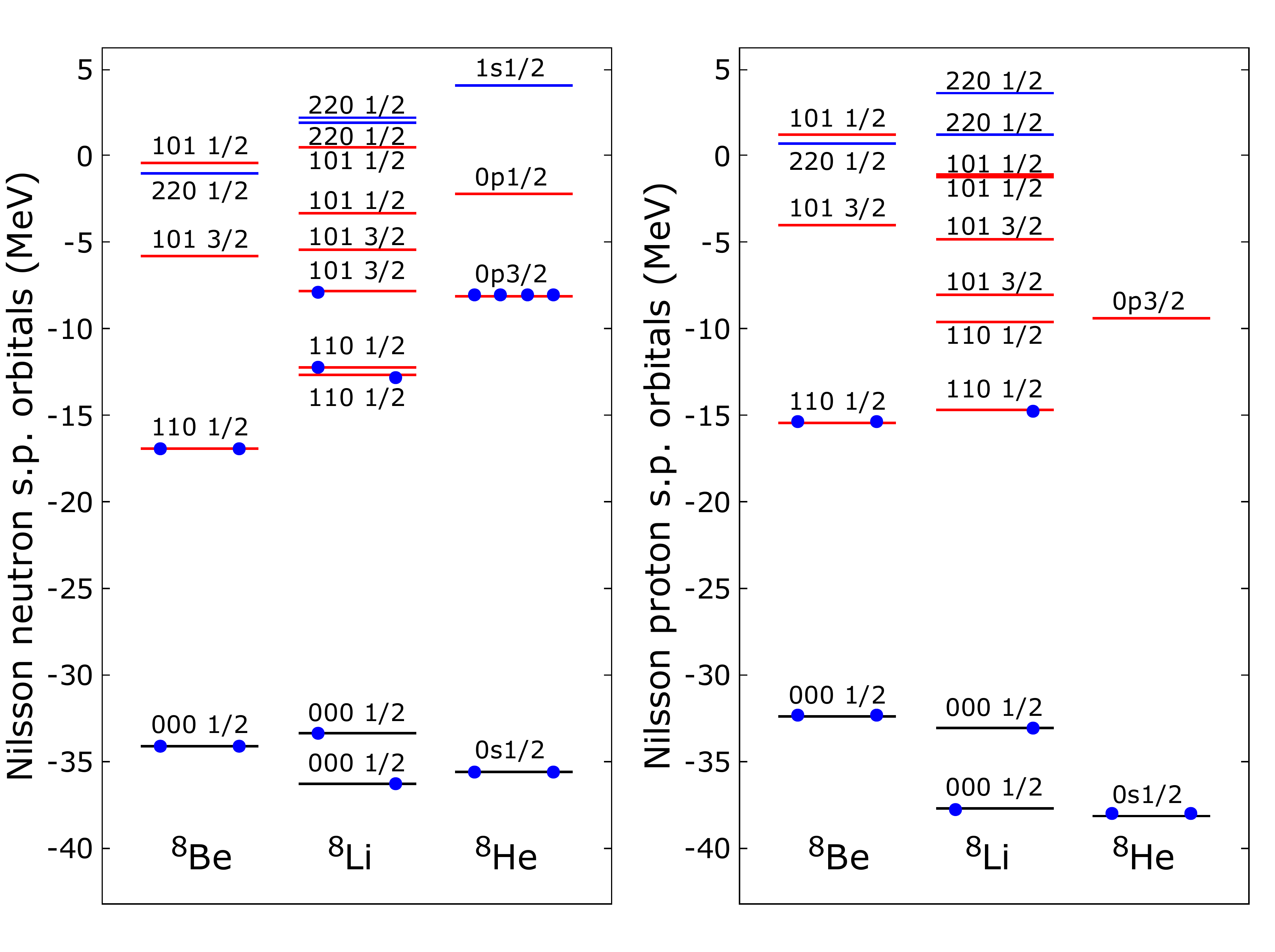}
\caption{(Colour online) Nilsson neutron (left) and proton (right) mean field single particle orbitals for SR ground state of $^8$Be, $^8$Li and $^8$He. 
The orbitals are labeled with approximate Nilsson quantum numbers. Dots indicate occupied levels and colours differentiate orbitals between parity-blocks.}
\label{fig:studniaA8}
\end{figure}

\subsection{Configuration space in $A$=8 nuclei}\label{conf:8}

Let us start the discussion by recalling the strategy of building configuration space.  As already 
discussed in Sect.~\ref{sec:ncci}, we start by calculating self-consistently the HF g.s. configuration. The s.p. Nilsson levels of both signatures (the signature symmetry is 
superimposed on our HF solutions) in the g.s. are used next as a guide to construct excited configurations. In the first place we include 
all relevant 1p-1h configurations. If needed, we extend the configuration space by adding low-lying 2p-2h configurations etc.

The neutron and proton s.p. Nilsson levels calculated for the ground states of $^8$Be, $^8$Li, and $^8$He are shown in Fig.~\ref{fig:studniaA8}. 
Note that the g.s. configurations of $^8$Be, $^8$Li are well deformed while the g.s. of $^8$He is spherical.
The split of s.p. level energies in $^{8}$Li is due to breaking of the time-reversal symmetry. 
In the case of  $N=Z$ nucleus $^{8}$Be, we built the space by taking into account the g.s. configuration. Next, we attempt to compute 
all four possible (aligned and anti-aligned) neutron 1p-1h excitations among the available 
$|N$=$1 n_z \Lambda\, \Omega\, \pm\rangle $ Nilsson states, where $\pm$ refers to the signature quantum number $r=\pm i$. 
It appears, however, that one of them,  the anti-aligned excitation to the first Nilsson s.p orbital $| 101\, 3/2 \rangle $, does not converge. 
Eventually, in an attempt to cover the missing correlations from the s.p orbital $| 101\, 3/2 \rangle $,  the configuration space consisting the g.s. and 
three $\nu p - \nu h$  is extended by adding three lowest 2p-2h excitations.

In the semi-magic nucleus $^{8}$He we include in the model space the g.s. and four 1p-1h neutron excitations.

In odd-odd $^{8}$Li, we compute first the aligned and anti-aligned g.s. configurations. Next, keeping the two neutrons paired in the lowest available 
signature reversed Nilsson states, we calculate several possible excited $|\nu\rangle \otimes |\pi \rangle$ configurations by distributing the unpaired 
proton and unpaired neutron over the available s.p. states.  Eventually, we break the neutron pair and attempt to compute fully unpaired configurations.  
These configurations are highly excited and difficult to converge. We were able to converge two such low-$K$ axial configurations. As it will be shown 
below, in Sect.~\ref{sec:A8}, they do not influence the low-energy part of the spectrum but have quite significant impact on the GT resonance.

\begin{table}[tbh]
\caption{Mean-field self-consistent configurations in $^{8}$He, $^{8}$Be, and
$^{8}$Li.  Configurations are ordered according to their excitation energies (index $i$) and 
labeled by the asymptotic Nilsson quantum numbers and the signature of unpaired valence 
particles and holes.  Last four columns list their properties including
HF energy in MeV,  quadrupole deformation parameters $\beta_2$ and $\gamma$, 
and the total alignment $\langle j \rangle$  and its orientation in the intrinsic frame, respectively. }
\label{tab:confA8}
\renewcommand{\arraystretch}{1.3}
\begin{ruledtabular}
\begin{tabular}{cccrrr}
$i$ & $|^{8}$He$;\, \varphi_i\rangle$        &    $E_{\rm HF}$  &   $\beta_2$   &  $\gamma$   & $\langle j \rangle$ \\
\hline
 1  &  $\nu p_{3/2}\otimes \pi s_{1/2}$  &$-$37.26   &   0       &     0$^\circ$   & 0\\
 2  &  $\ket{\nu 101\, 3/2\, -}^{-1} \otimes \ket{\nu 101\, 1/2\, +}^1$  &$-$32.47   &   0.14       &    0$^\circ$   &  $2_Z$\\
 3  &  $\ket{\nu 101\, 3/2\, +}^{-1} \otimes \ket{\nu 101\, 1/2\, -}^1$  &$-$30.81   &   0.03       &     60$^\circ$   &  $1_Y$\\
 4  &  $\ket{\nu 110\, 1/2\, +}^{-1} \otimes \ket{\nu 101\, 1/2\, +}^1$  &$-$30.04   &   0.03       &    60$^\circ$   &  $0_Y$\\
 5  &   $\ket{\nu 110\, 1/2\, +}^{-1} \otimes \ket{\nu 101\, 1/2\, -}^1$  &$-$29.13   &   0.02       &    0$^\circ$   &   $1_Z$\\
\hline
$i$ & $|^{8}$Be$;\, \varphi_i\rangle$        &     $E_{\rm HF}$  &   $\beta_2$   &  $\gamma$   & $\langle j \rangle$    \\
\hline
1  &  $\ket{\nu 110\, 1/2}^2 \otimes \ket{\pi 110\, 1/2}^2 $  &$-$48.66   &   0.68       &     0$^\circ$   & $0_Z$\\
2  &  $\ket{\nu 110\, 1/2\, -}^{-1} \otimes \ket{\nu 101\, 3/2\, +}^1$  &$-$38.87   &   0.40       &    0$^\circ$   &  $1_Z$\\
3  &  $\ket{\nu 110\, 1/2\, -}^{-1} \otimes \ket{\nu 101\, 1/2\, +}^1$  &$-$34.08   &   0.39       &     0$^\circ$   &  $1_Y$\\
4  &  $\ket{\nu 110\, 1/2\, +}^{-1} \otimes \ket{\nu 101\, 1/2\, +}^1$  &$-$31.63   &   0.27       &    3$^\circ$   &  $0.7_Z$\\
5  &   $\ket{\nu 110\, 1/2\, +}^{-1} \otimes \ket{\nu 101\, 3/2\, +}^1$  &$-$36.81   &   0.20       &    60$^\circ$   &   $0_Z$\\
 \mbox{}  &   $\ket{\pi 110\, 1/2+}^{-1} \otimes \ket{\pi 101\, 3/2\, +}^1$  & \mbox{}   &   \mbox{}       &     \mbox{}  & \mbox{}\\
6  &   $\ket{\nu 110\, 1/2}^{-2} \otimes \ket{\nu 101\, 3/2}^2$  &$-$35.74   &   0.11       &    5$^\circ$   &   $0_Z$\\

 7  &   $\ket{\nu 110\, 1/2\, +}^{-1} \otimes \ket{\nu 101\, 3/2\, +}^1$  &$-$34.28   &   0.12       &    0$^\circ$   &   $2_Z$\\
 \mbox{}  &   $\ket{\pi 110\, 1/2\, +}^{-1} \otimes \ket{\pi 101\, 3/2\, +}^1$  & \mbox{}   &   \mbox{}       &     \mbox{}  & \mbox{}\\

\hline
$i$ & $|^{8}$Li$;\, \varphi_i\rangle$        &     $E_{\rm HF}$  &   $\beta_2$   &  $\gamma$   & $\langle j \rangle$   \\
\hline
 1  &  $\ket{ \nu 101 \, 3/2\, +} \otimes \ket{ \pi 110 \, 1/2\, +}$   &$-$39.08  &   0.38       &     0$^\circ$  & $1_Z$\\
 2  &  $\ket{ \nu 101 \, 3/2\, +} \otimes \ket{ \pi 110 \, 1/2\, -}$   &$-$39.03  &   0.36       &     0$^\circ$  & $2_Z$\\
 3  & $\ket{ \nu 101 \, 1/2\, +} \otimes \ket{ \pi 110 \, 1/2\, +}$   &$-$34.04  &   0.36       &     
0$^\circ$ & $1_Z$\\
 4  & $\ket{ \nu 101 \, 1/2\, -} \otimes \ket{ \pi 110 \, 1/2+}$    &$-$33.44  &   0.35       &     0$^\circ$ & $0_Z$\\
 5   & $\ket{ \nu 110 \, 1/2\, +} \otimes \ket{ \pi 110 \, 1/2-}$  &$-$36.51  &   0.07       &    
60$^\circ$  & $0_Z$\\
 6  & $\ket{ \nu 101 \, 3/2\, +} \otimes \ket{ \pi 101 \, 3/2+}$   &$-$35.68  &   0.03       &     
0$^\circ$  & $0_Y$\\
 7   & $\ket{ \nu 101 \, 3/2\, +} \otimes \ket{ \pi 101 \, 1/2\, -}$  &$-$32.34  &   0.12       &    
0$^\circ$  & $2_Z$\\
 8  & $\ket{ \nu 101 \, 1/2\, +} \otimes \ket{ \pi 110 \, 1/2\, +}$    &$-$31.19  &   0.06       &     60$^\circ$ & $1_Z$\\
 9   & $\ket{ \nu 101 \, 3/2\, +} \otimes \ket{ \nu 110 \, 1/2\, +}  $  &$-$29.25  &   0.04       &    
60$^\circ$  & $0_Y$\\
\mbox{} & $ \otimes \ket{ \nu 101 \, 1/2\, -}   \otimes \ket{ \pi 101 \, 3/2\, -}$  & \mbox{} & \mbox{} &
\mbox{} & \mbox{}  \\  
 10   & $\ket{ \nu 101 \, 3/2\, +} \otimes \ket{ \nu 110 \, 1/2\, +}  $  &$-$29.06  &   0.07       &    
60$^\circ$  & $1_Y$\\
\mbox{} & $ \otimes \ket{ \nu 101 \, 1/2\, +}   \otimes \ket{ \pi 101 \, 3/2\, -}$  & \mbox{} & \mbox{} &
\mbox{} & \mbox{}  \\  

\end{tabular}
\end{ruledtabular}
\end{table}

All configurations included in the configuration spaces of $^8$Be, $^8$Li, and $^8$He are listed in Table~\ref{tab:confA8}. The configurations
are labeled by means of the Nilsson and signature quantum numbers, $|N n_z \Lambda\, \Omega\, \pm \rangle$,
pertaining to the unpaired valence particles. Table also includes quadrupole deformation parameters $\beta_2$ and $\gamma$
for each configuration. A value of $\gamma \neq 0^\circ$ or $\gamma \neq 60^\circ$ indicates a tri-axial configuration.

We use the Nilsson quantum numbers to label not only deformed but also near-spherical configurations. This is partly justified 
since  some of these configurations, in particular those in  $^{8}$Li, exhibit very peculiar isovector shape effects. 
For example, the near-spherical configuration 6 in $^{8}$Li is a superposition of prolate (oblate) density distribution 
of neutrons (protons), respectively, while  the configurations 5 and 8, are superpositions of oblate (prolate) density distributions of neutrons (protons), 
respectively. The near-spherical configurations 7, 9 and 10, on the other hand, are built of near-spherical density distribution
of protons 7 (neutrons 9 and 10) and deformed density distribution of neutrons (protons), respectively.  
Note, that these isovector shape effects may lead to different $\Omega$-ordering of the neutron and proton s.p. levels.

\begin{figure}[htb]
\centering
\includegraphics[width=1.0\columnwidth]{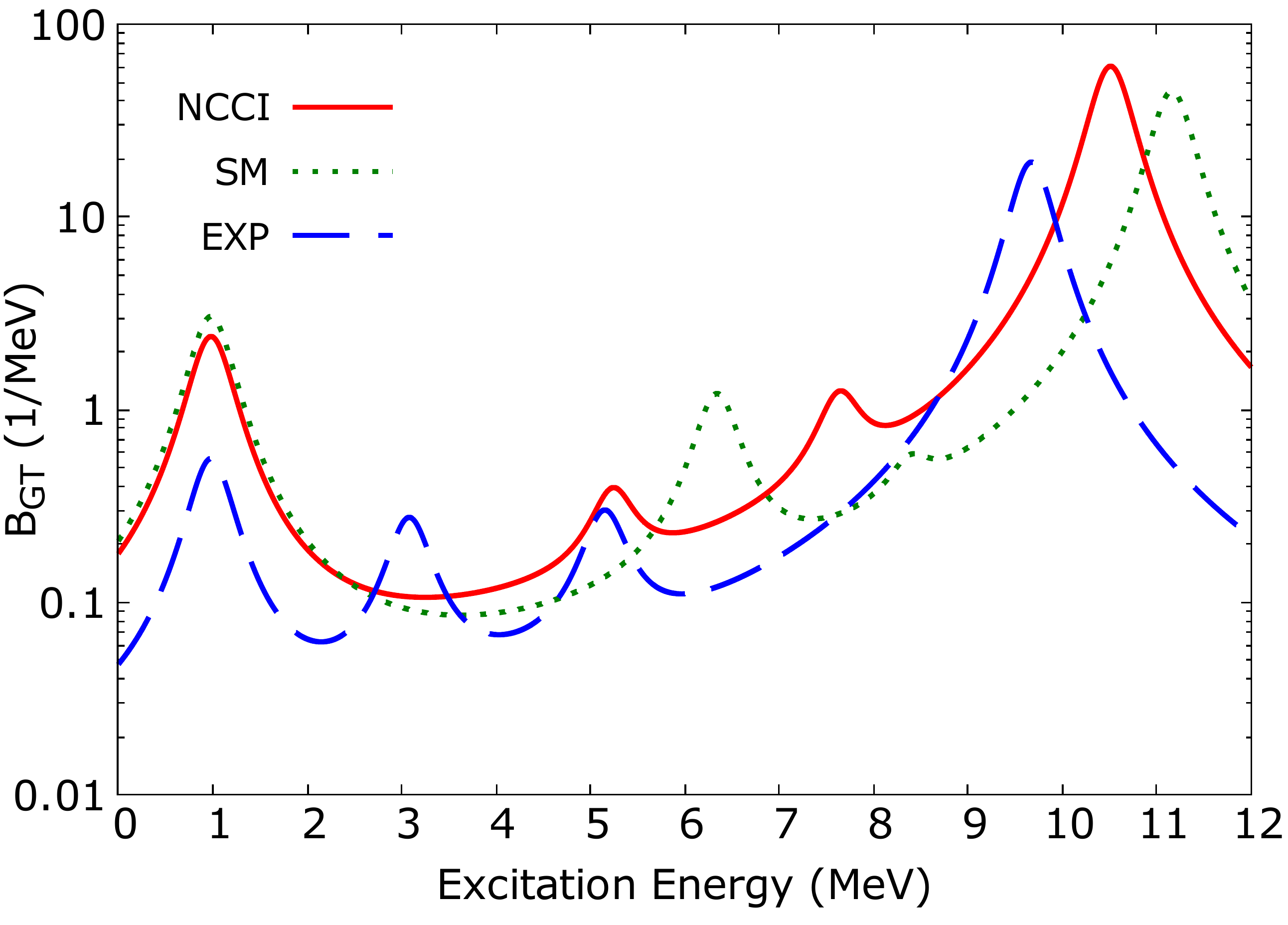}
\caption{(Colour online) The GT strength distribution in $1^+$ states of $^{8}$Li in the logarithmic scale smoothed with Lorentzian function with half-width of $\Gamma=0.5$MeV. The dashed curve represents experimental data obtained by means of the R-matrix theory in Ref.~\cite{(Bar88),(Til04)}.  The dotted line marks the shell-model input to the R-matrix~\cite{(Bar88)} 
calculated using the $p$-shell residual interaction proposed by Kumar in Ref.~\cite{(Kum74)}. The continuous line labels the DFT-NCCI calculations. See text for discussion.}
\label{fig:GTSDli}
\end{figure}

\subsection{GT strength distribution for $^8$He to $^8$Li decay}

Beta decay from the $0^{+}$ g.s. of $^{8}$He populates four $1^{+}$ states in $^{8}$Li within the experimental $Q_{\beta}$ energy window. 
Except for the lowest $1^+$ state, the remaining $1^+$ states may decay through different particle emission channels. It makes therefore both 
the energy and B$_{\textrm{GT}}$ of decaying-states extremely difficult to determine experimentally. In fact, the so called experimental determination
of $\beta$-decay properties of $^{8}$He,  is based on multi-parameter R-matrix formalism. The initial values of the R-matrix parameters are  
taken from  shell-model calculations.  These parameters are varied next to best fit the available data on half-life, branching ratios 
and energy spectra of beta-delayed  particles~\cite{(Bar88),(Bor93),(Bar96)}. The  inclusion of the particle emission channels reduces the 
experimental B$_{\textrm{GT}}$ to the resonant $1^{+}$ states in $^{8}$Li and shifts their energies (centroids) as compared to 
the initial shell-model values with the largest impact on the GT resonance - the fourth $1^+$ state - which can decay through both the neutron and 
triton emission, as shown in Fig.~\ref{fig:GTSDli}.

In Fig.~\ref{fig:GTSDli} we present GT strength function for the decay of $0^{+}$ g.s. of $^{8}$He, smoothed with the 
Lorentzian distribution of half-width of $\Gamma=0.5$\,MeV.  The peaks in the distribution reflect excitation energies of the GT-populated
$1^{+}$ states in $^{8}$Li. Distribution is normalized with respect to the first $1^+$ state which is bound.

The DFT-NCCI model predicts a $2^+$ g.s. in $^{8}$Li at the energy of $-41.9$\,MeV, which is only  $\sim$0.6\,MeV below the experimental value. 
The resonant peak in the DFT-NCCI spectrum is shifted by $\sim$1\,MeV towards higher energies as compared to experimental data, whereas
the second and third peaks are roughly 2\,MeV higher than the experiment. 
The height of the peaks is overestimated, in particular, for the GT resonance.
Naturally, such a big difference cannot be explained solely by the quenching factor, which is {\it a fortiori\/} expected to be close to unity in light nuclei. 
The discrepancy is mostly due to lack of the coupling to particle-emission channels in the DFT-NCCI. In this respect, our results should be considered 
as an input to the R-matrix and compared directly to the shell-model input to the R-matrix. Such a comparison shows, see  Fig.~\ref{fig:GTSDli}, that the results 
on the GT strength distribution are very similar. This is benefiting for us since  our calculations are free from any adjustable parameter at variance to the shell-model 
results of Ref.~\cite{(Kum74)}. Part of the discrepancy may be also due to the  three-nucleon forces which, in {\it ab initio\/} NCCI calculation may become prominent, as was 
found in the beta decay of $^{14}$C~\cite{(Mar11)}.

The DFT-NCCI approach allows for rather unique analysis of the GT strength distribution in terms of HF configurations
which are the primary building blocks of the model. This is particularly useful in deformed nuclei where HF
configurations, corresponding to certain p-h excitations, can be conveniently and rather intuitively labeled by 
Nilsson quantum numbers. The content of the $n$-th HF configuration in the $k$-th DFT-NCCI state of a given 
$I$ and $T_z$, see Eq.~\ref{eq:nccistate}, that corresponds to the $k$-th peak in the spectrum is 
given by the following formula:
\begin{eqnarray}
P (\varphi_n ) \equiv \sum_i |  ^{(i)}\langle\varphi_n ; IM; T_z  | \psi_{\textrm{NCCI}}^{k;IM;T_z}\rangle |^2 \nonumber \\
&\phantom{=}&\hspace{-2.7in} 
=\frac{1}{\mathcal{N}^{(k)}_{IM;T_z}}   \sum_i|\sum_{jl} c_{jl}^{(k)}\,  
{^{(i)}\langle\varphi_n; IM; T_z |\varphi_j; IM; T_z}\rangle^{(l)}|^2 \,. \label{eq:content}
\end{eqnarray}

\begin{figure}[htb]
\centering
\includegraphics[width=1.0\columnwidth]{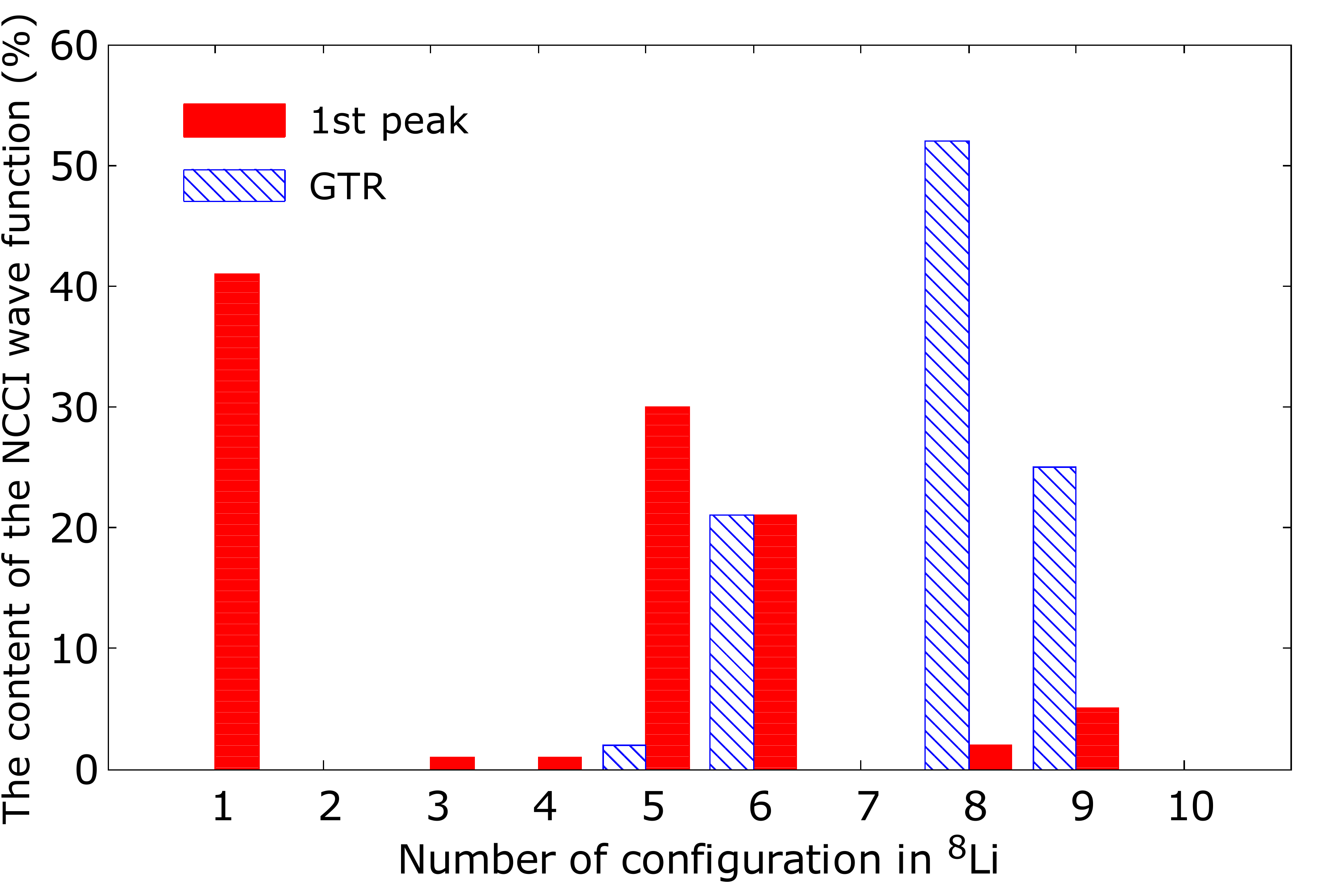}
\caption{(Colour online) Decomposition of the wave functions of the first and fourth (GT resonance) $1^+$ states 
in $^{8}$Li in terms of the HF configurations, $\varphi_n$, included in the space. An abscissa 
numbers the configurations  
in accordance with Table~\ref{tab:confA8}.}
\label{fig:GTGRli}
\end{figure}

Fig.~\ref{fig:GTGRli} shows a decomposition of the wave functions of first and fourth $1^+$ states in  $^{8}$Li in 
terms of the included HF configurations. These HF configurations are the same as listed in  Table~\ref{tab:confA8}. 
As shown in the figure, the first peak is a mixture of the very well deformed aligned ground state, $|^{8}$Li$;\varphi_{1}\rangle$, 
with two very weakly deformed proton excitations $i=5$ and $6$. The lowest proton-excited configuration corresponds to 
oblate shape. The second is proton-excited configuration corresponding to prolate shape.

The resonance is centered around weakly deformed oblate configuration, $|^{8}$Li$; \varphi_{8} \rangle$, corresponding to 1p-1h aligned excitation from 
$\ket{ \nu 101 \, 3/2  \, +}$ Nilsson level to its spin-orbit partner $\ket{\nu 101 \, 1/2  \, -}$ with drastic shape-change of neutron density. An admixture of broken-neutron-pair configuration 9, to the resonance is of the order of $25\%$. And finally, $20\%$ of a resonant peak comes from the lowest proton excitation to the $\ket{ \pi 101 \, 3/2  \, +}$ Nilsson level.

\subsection{GT strength distribution for $^8$Li to $^8$Be decay}

The  $^8$Be nucleus is a cluster composed of two $\alpha$ particles. Its molecular structure is characterized by 
very elongated  distribution of nuclear matter which is well accounted for by our mean-field calculation which
predicts a sudden increase of deformation in $^8$Be to $\beta_2=0.68$ as compared to its neighbours. It appears, however, that neither the 
HF nor the DFT-NCCI  can account for all correlations associated with the clustering. The g.s. energy calculated 
using the DFT-NCCI equals $-52.8$\,MeV, underestimating the experimental value by 3.7\,MeV. This should be compared to
the g.s. energy of $^{8}$Li which was overestimated only by 0.6\,MeV.

\begin{figure}[htb]
\centering
\includegraphics[width=\columnwidth]{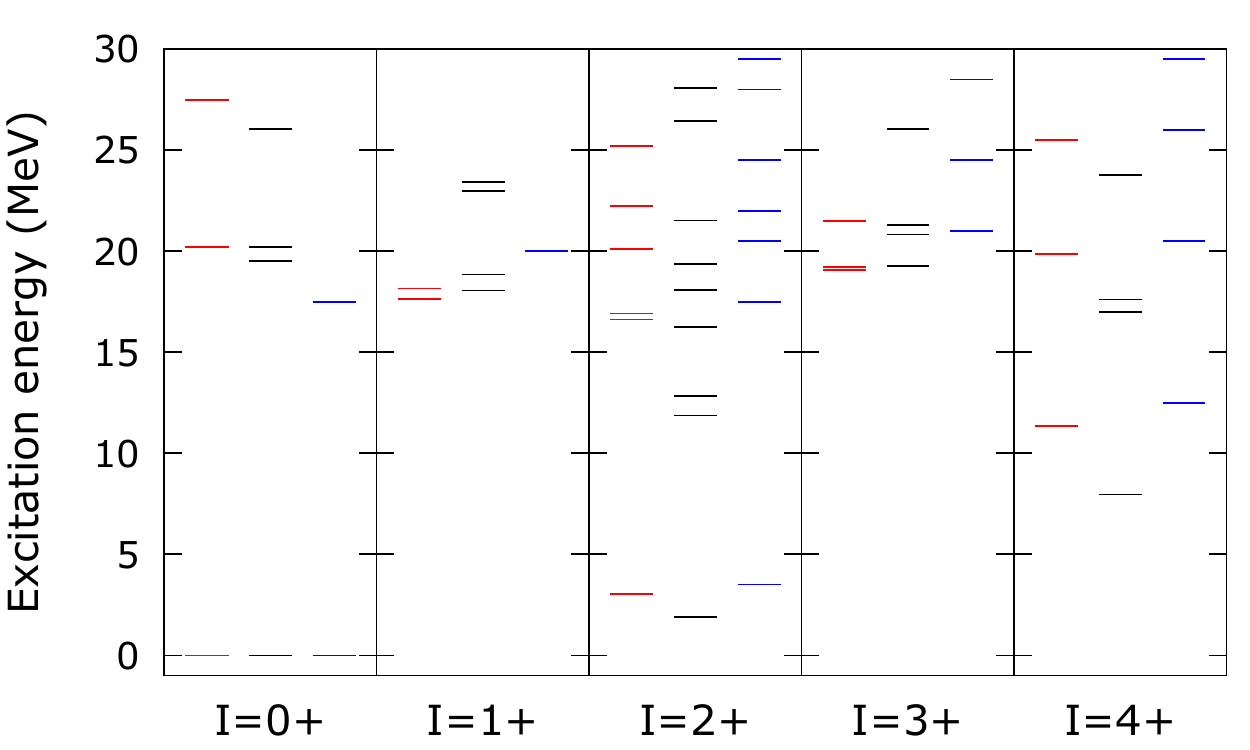}
\caption{(Color online) Low-spin states in $^8$Be below 30\,MeV. The panels show, counting from the left, the groups of levels having spins 
$I^\pi=0^+,1^+,2^+,3^+$, and $4^+$, respectively.  Each panel shows experimental (left), DFT-NCCI (center) and ab initio 
NCCI (right) spectra, each normalized with respect to its g.s. energy. 
}
\label{fig:8be-spektrum}
\end{figure}

The low-spin positive-parity levels in $^8$Be are shown in Fig.~\ref{fig:8be-spektrum}. 
Apart of experimental data and the results of DFT-NCCI, the figure includes also, for a sake of comparison, 
the results of {\it ab initio\/} NCCI calculations of Ref.~\cite{(Mar15)} with JISP16 interaction.  This
calculation predicts the g.s. at $-57.5$\,MeV i.e. roughly 1\,MeV below the experiment. 
A similar {\it ab initio\/} NCCI calculation in Ref.~\cite{(Cap15)} with NNLO chiral potential
results to under bounded g.s. energy.

As shown in the figure, our calculations reproduce relatively well odd-spin states.  The level of agreement is comparable, if not 
better, to the {\it ab initio\/} NCCI results. The calculated isospin doublet of $1^+$ states 
around $24$\,MeV may represent a doublet seen experimentally at 23\,MeV. Spins for this doublet has not yet been assigned. 
Even-spin states, on the other hand, are systematically overbound.  
The lowest 2$^+$ and 4$^+$ states are interpreted 
as a members of a rotational band built atop of the 0$^+$ g.s. Their empirical excitation energy ratio,  $R_{4/2}\equiv {E_{4_1^+}}/{E_{2_1^+}}$,
equals  3.75 and thus belongs to the largest over the entire nuclear chart. Our model  captures quite well the ratio giving  $R_{4/2}=3.77$. 
This means that our DFT-rooted calculation reproduces change of the moment of inertia along the band well but strongly overestimates its magnitude.  

Too strong a quadrupole collectivity ($2_1^+$ and $4_1^+$  are to low in energy) and missing correlations in the calculated g.s. are well seen in the GT transition strengths of
$(^8$Li, $2_{\textrm{g.s.}}^+)$ to $(^8$Be, $2^+_i)$ decays. The DFT-NCCI results and experimental data are compared in Table~\ref{table:GTSDbe}. The DFT-NCCI results are 
calculated within 1p-1h configuration space. Inclusion of 2p-2h configurations has only a marginal impact on the results.
The transition strength to the $2^+_1$ state is clearly overestimated by our model,
in contrary to the transition strength to the $2^+_2$ resonance, which seems to be 
underestimated. One should bear in mind, however, that the empirical strength to the 
resonance is uncertain and can be affected by the close-lying $2^+_3$, $T=1$ state. 

\begin{table}[htb]
\caption{Experimental and theoretical excitation energies of  the three lowest $2^+$ states in $^8$Be and the corresponding $\log ft$ values.  Experimental data are taken from Ref.~\cite{(Ajz84)}. 
}
\label{table:GTSDbe}
\begin{ruledtabular}
\begin{tabular}{rrrrrrr}
 & \multicolumn{2}{c}{Experiment} & \multicolumn{2}{c}{DFT-NCCI} \\
State & $E$ (MeV) & $\log ft$ & $E$ (MeV) & $\log ft$ \\
\hline
2$_{1}^{+}$ $T=0$ & 3.030  & 5.36 & 2.698 & 4.74 \\
2$_{2}^{+}$ $T=0$ & 16.626 & 2.93 & 11.869 & 3.54 \\
2$_{3}^{+}$ $T=1$ & 16.922 & $-$   & 12.812 & 4.13 \\
\end{tabular}
\end{ruledtabular}
\end{table}

\subsection{Gamow-Teller sum rule -- configuration and model space dependence}

The Gamow-Teller sum rule (GTSR) is commonly considered as a convenient 
indicator of the completeness of a model space. Under the assumption of completeness the GTSR reads as follows
\begin{eqnarray}
\frac{1}{g_{\rm A}^2}\sum_{f} \Big{[}B_{\text{GT}}^-(I_i^\pi\to I_f^\pi) \nonumber \\ 
&\phantom{-}&\hspace{-1.0in} -B_{\text{GT}}^+(I_i^\pi\to I_f^\pi)\Big{]}=3(N-Z) \,,
\end{eqnarray} 
where the sum extends over all final states  $I_f=I_i+k$ with $k=0,\pm 1$. The strength is defined as
\begin{equation}
B_{\text{GT}}^{\pm}(I_i^\pi\to I_f^\pi)
= g_{\text{A}}^2 \frac{ |M^{\pm}_{\text{GT}}|^2}{2I_i+1}
\end{equation}
where $M_{\textrm{GT}}^{\pm}$ stands for reduced matrix element for the Gamow--Teller one-body operator.

  In this section we shall discuss the  
GTSR in $^8$Li calculated within the DFT-NCCI with a particular emphasis on its  dependence on the configuration 
and model spaces. The configurations (HF solutions) are numbered and labeled as in Table~\ref{tab:confA8}.
The model space, on the other hand, is spanned by so-called {\it natural states\/}. 
These are linearly independent linear combinations of projected states Eq.(~\ref{eq:nccistate}) having eigenvalues of the norm 
matrix, $n_i $,  larger than a certain externally provided cut-off parameter  $\varepsilon $.

\begin{figure}[htb]
\centering
\includegraphics[width=1\columnwidth]{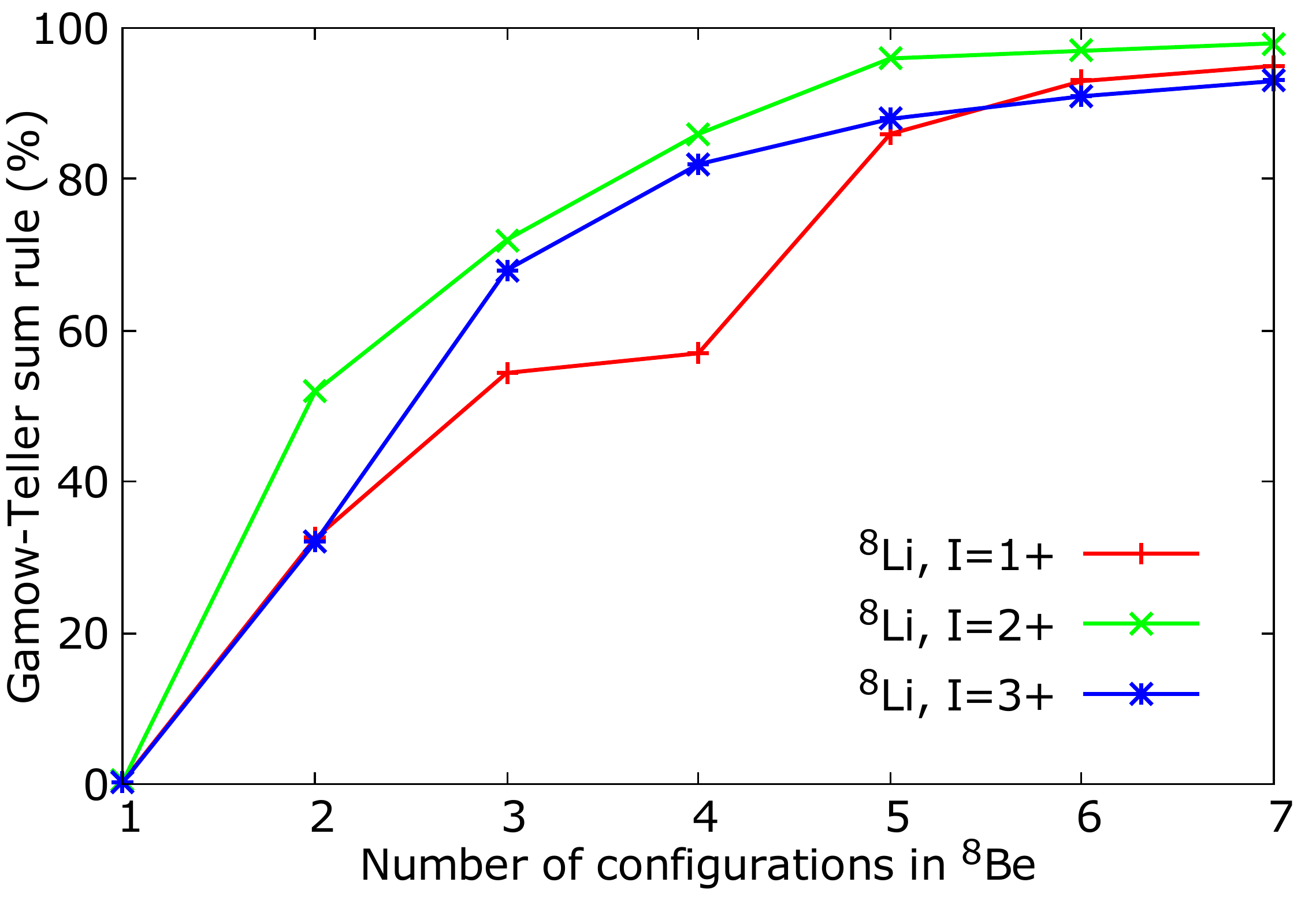}
\caption{(Colour online) GTSRs for  the lowest $1_1^+$, $2_1^+$  and $3_1^+$  initial state in $^8$Li against the number of configurations included in $^8$Be. The configurations are listed in Table~\ref{tab:confA8}.}
\label{fig:GTSRbe}
\end{figure}

In Fig.~\ref{fig:GTSRbe} we show the saturation of GTSRs for the $^8$Li $1_1^+$, $2_1^+$ and $3_1^+$ initial states
versus a number of configurations used in $^8$Be final state.
In the calculations, the $B^+_{\text{GT}}$ was kept fixed at a value calculated using the entire 1p-1h configuration space in 
$^8$He and $^8$Li, see Table~\ref{tab:confA8}.  It is benefiting to observe that already with five configurations 
in $^8$Be, the calculated GTSR reaches a level of 90\%. The remaining 2p-2h  provide circa $5\%$ of the strength. 
It is interesting to note also that the unconverged 1p-1h configuration involving $\ket{101\, 3/2}$ Nilsson orbit can be effectively 
replaced by 2p-2h excitations to this orbital.

\begin{figure}[htb]
\centering
\includegraphics[width=1.0\columnwidth]{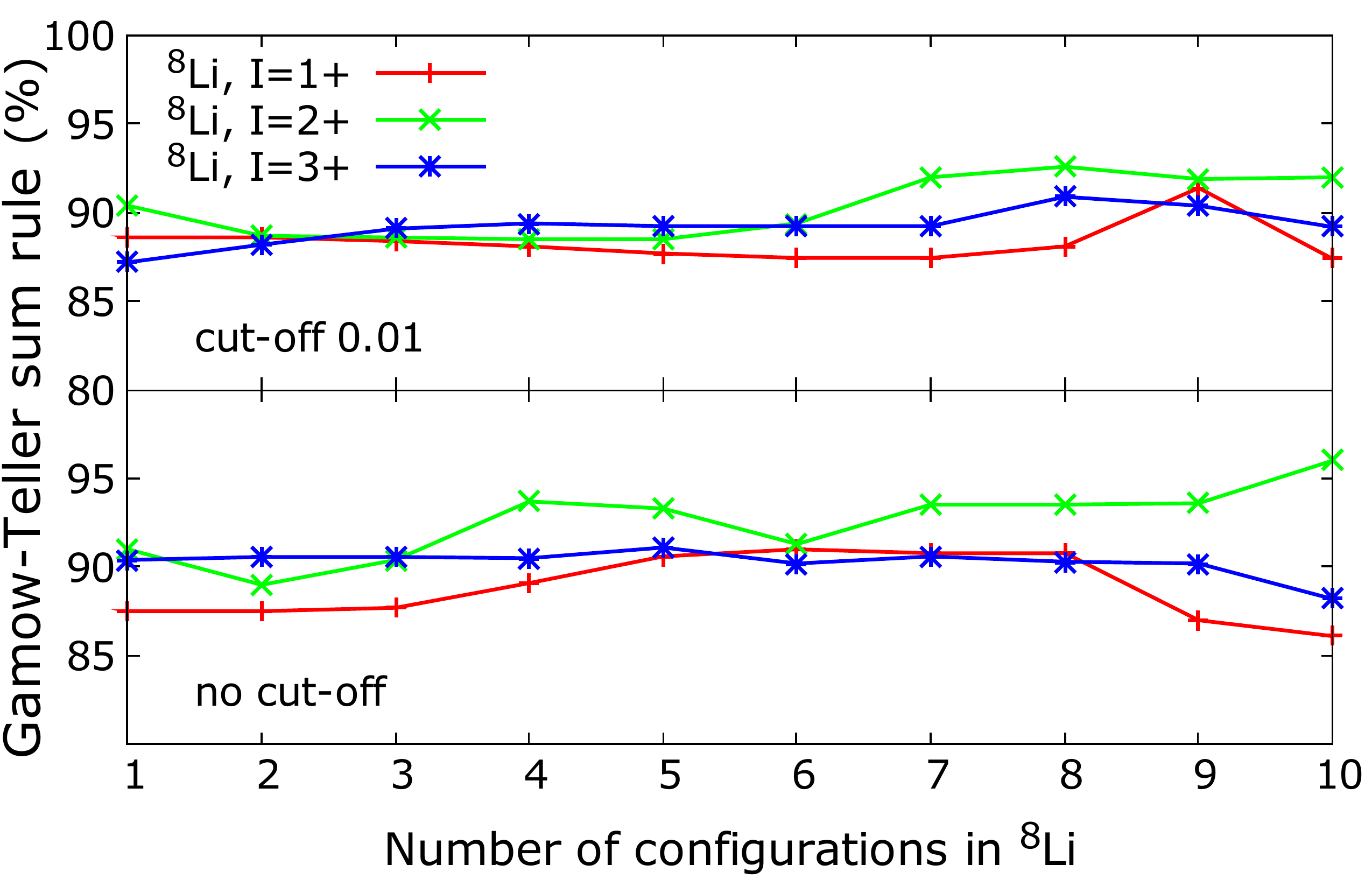}
\caption{(Colour online) GTSRs for the first $1_1^+$, $2_1^+$  and $3_1^+$  initial state in $^8$Li against the number of configurations included in $^8$Li. Bottom panel shows the GTSR calculated without the cut-off.
Upper panel represents the results obtained for $\varepsilon \approx 0.01$. See text for further details.}
\label{fig:GTSRli}
\end{figure}

Fig.~\ref{fig:GTSRli} shows the GTSRs in $^8$Li and its sensitivity  with respect to the cut-off parameter $\varepsilon$. 
The calculations are performed for $2_1^+$, $1_1^+$,  and $3_1^+$ states in $^8$Li. 
In the calculations we fix the number of configurations in $^8$Be and $^8$He, by taking five configurations 
in each nucleus, and add configurations in $^8$Li to study the saturation of the GTSR. 
In the bottom panel we present the results without any cut-off. With a single g.s. configuration in $^8$Li  we 
reach $\sim 90\%$ of the sum rule irrespectively on spin.  The GTSR value 
does not change much with increasing number of included configurations in  $^8$Li. The reason comes from the fact that within our framework lowest-lying $2^+_1$, $1^+_1$ and $3^+_1$ states have their origins in the g.s. configuration, which captures most of the important correlations.

\begin{figure}[htb]
\centering
\includegraphics[width=1.0\columnwidth]{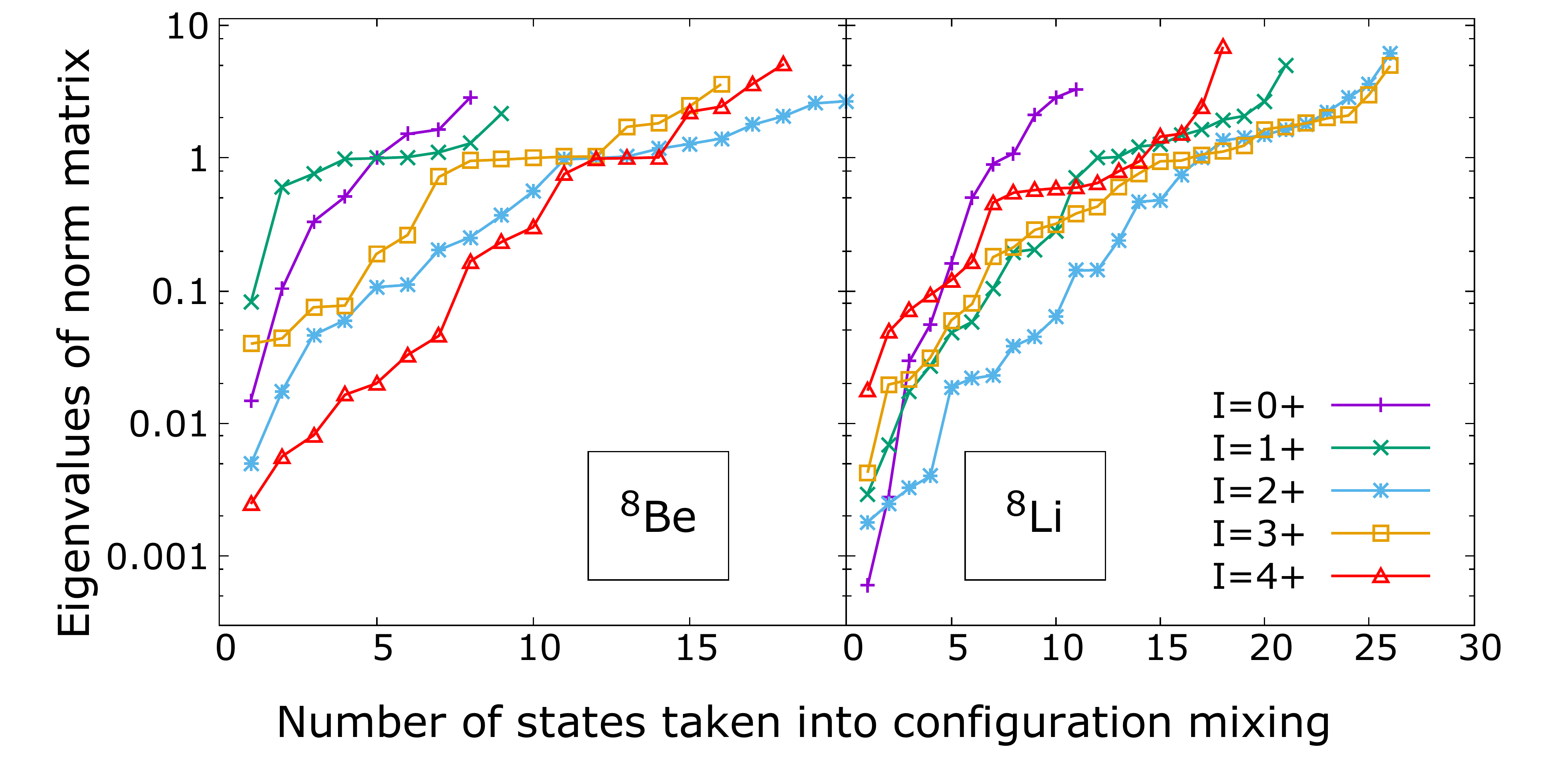}
\caption{(Colour online) Eigenvalues of the norm matrix for $0^+$, $1^+$, $2^+$, $3^+$, and 
$4^+$ states in $^{8}$Be (left) and in $^{8}$Li (right). The eigenvalues are plotted in ascending order.}
\label{fig:normy}
\end{figure}

In many cases the {\it natural states\/}, corresponding to small eigenvalues of the norm matrix, lead to 
instabilities in DFT-NCCI calculation. The instabilities can be controlled to some extent by applying the appropriate 
cut-off parameter  $\varepsilon$. The choice of the cut-off parameter is, however, not unique. Typically, its 
value is correlated with discontinuities (or jumps) seen in the eigenvalues of the norm matrix plotted in ascending (or descending) 
order.  In $^8$Li, see Fig.~\ref{fig:normy},  the most natural choice is $\varepsilon \approx 0.01$.  This choice, as shown
in Fig.~\ref{fig:GTSRli}, has almost no impact on the GTSR. With increasing $\varepsilon$, 
more physical states are being removed, which, in turn,  gives a rise to large variations of the GTSR versus number of configurations.   

\section{Gamow-Teller strength distribution in the sd-midshell nucleus $^{24}$Mg }\label{sec:sdSD}

In this section we present the DFT-NCCI results for the Gamow-Teller strength distribution (GTSD) in $^{24}$Mg following the g.s. beta decay of $^{24}$Al ($I_{\text{g.s.}}^\pi =4^+$).
For similar analysis of the GTSD in the neighbouring nucleus $^{20}$Ne we refer reader to our conference publication~\cite{(Kon17)}.

\begin{figure}[htb]
\centering
\includegraphics[width=0.6\columnwidth]{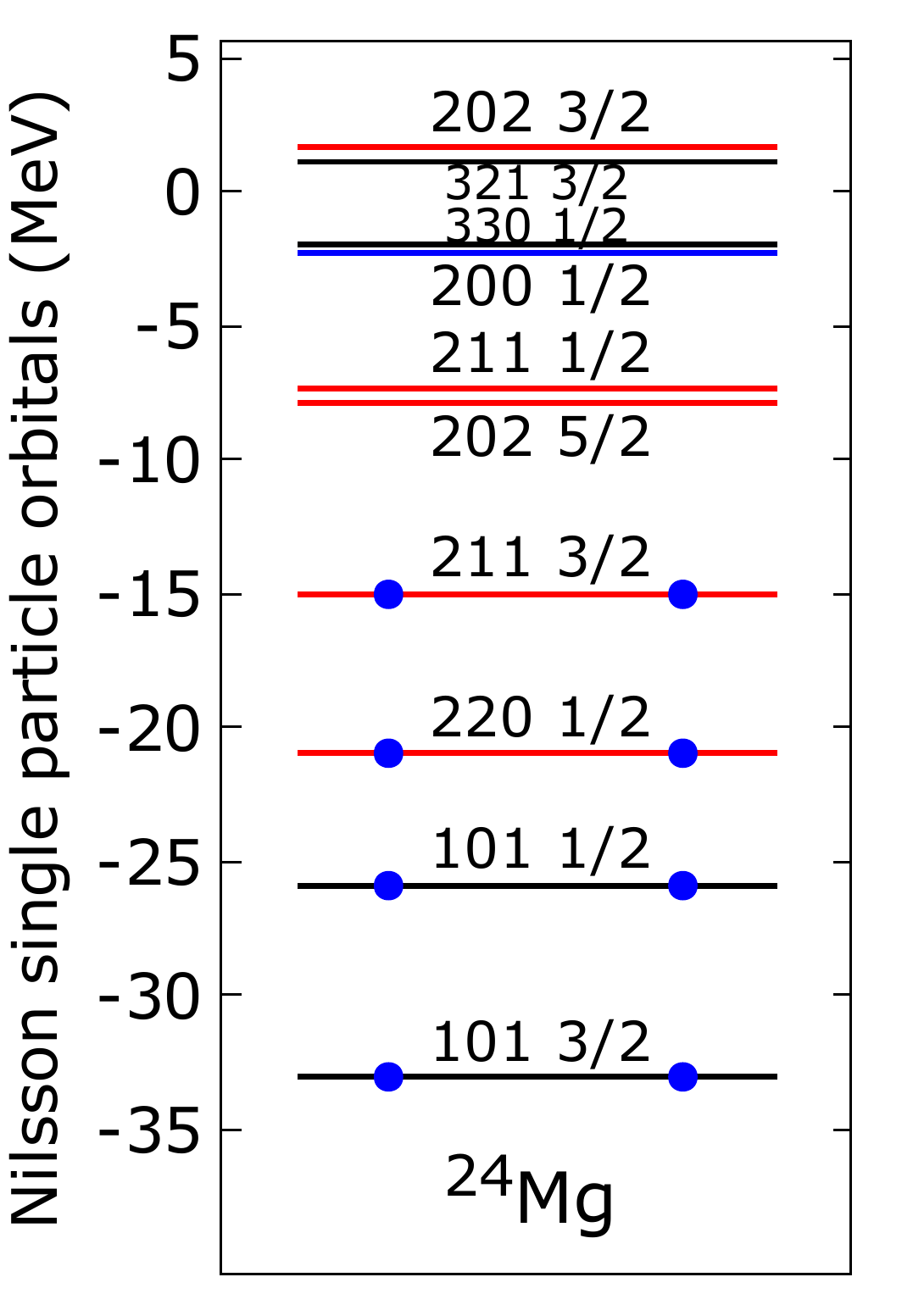}
\caption{(Colour online) Neutron mean field s.p. levels in the g.s. of $^{24}$Mg.  
The levels are labeled using the asymptotic Nilsson quantum numbers. Dots indicate occupied levels.}
\label{fig:studnia}
\end{figure}

Within the conventional spherical shell-model terminology, $^{24}$Mg is a $sd$-shell nucleus having eight valence particles. Mean-field calculations, on the other hand, predict $^{24}$Mg to be well deformed system. Hence, the DFT-NCCI configuration space is built by promoting particles among the deformed s.p. Nilsson levels as shown in Fig.~\ref{fig:studnia}. 

In order to facilitate 
the discussion below, let us recall that the Nilsson levels $\ket{ 220\, 1/2}$,  $\ket{ 211\, 3/2}$, and $\ket{ 202\, 5/2}$  originate 
from the spherical $d_{5/2}$ sub-shell, the level  $\ket{ 200\, 1/2}$ comes from the spherical $s_{1/2}$ sub-shell, and  
the levels $\ket{ 211\, 1/2}$ and $\ket{ 202\, 3/2}$ originate from the spherical  $d_{3/2}$ sub-shell. Moreover, the levels 
$\ket{ 200\, 1/2}$ and $\ket{ 211\, 1/2}$ are predicted to mix through the quadrupole field when intrinsic deformation is $\beta_2 \sim 0.1-0.3$, 
see for example the Nilsson diagram in Ref.~\cite{(RS80)}. 

\begin{table}[!h]
\caption{The list of configurations in $^{24}$Mg, labeled by
the index $i$ and asymptotic Nilsson quantum numbers of excited p-h states. Listed are
also the HF binding energy $E_{\rm HF}$ in MeV, excitation energy $\Delta E$ in MeV,
quadrupole deformation parameters $\beta_2$, and the total alignment $K$ together with its orientation in the intrinsic frame.}
\label{tab:confsd}
\renewcommand{\arraystretch}{1.3}
\begin{ruledtabular}
\begin{tabular}{cccrrr}
$i$ & $|^{24}$Mg$;i\rangle$        &     $E_{\rm HF}$  & $\Delta E$     &  $\beta_2$   & $K$   \\
\hline
1  &   g.s.  &$-$194.33   & 0 &  0.42   &   $0$\\
2  &  $\ket{\nu 211\, 3/2 -}^{-1} \otimes \ket{\nu 202\, 5/2 -}^1$  &$-$187.92   & 6.41 &   0.34   &  $1_Z$\\
3  &  $\ket{\nu 211\, 3/2 +}^{-1} \otimes \ket{\nu 202\, 5/2 -}^1$  &$-$187.25   & 7.08 &   0.34  &  $4_Z$\\
4  &  $\ket{\nu 211\, 3/2 +}^{-1} \otimes \ket{\nu 211\, 1/2 -}^1$  &$-$187.46   & 6.87  & 0.43   &  $2_Z$\\
5  &   $\ket{\nu 211\, 3/2-}^{-1} \otimes \ket{\nu 211\, 1/2-}^1$  &$-$184.89   & 9.44 &  0.40   &   $1_Z$\\
6  &   $\ket{\nu 220\, 1/2-}^{-1} \otimes \ket{\nu 202\, 5/2-}^1$  &$-$183.34   & 10.99 &  0.24    &   $2_Z$\\
 7  &   $\ket{\nu 220\, 1/2+}^{-1} \otimes \ket{\nu 202\, 5/2-}^1$  &$-$183.27   & 11.06  & 0.23  &   $3_Z$\\
8  &  $\ket{\nu 211\, 3/2 +}^{-1} \otimes \ket{\nu 200\, 1/2 +}^1$  &$-$181.79   & 12.54 &   0.36   &  $1_Z$\\
9  &  $\ket{\nu 211\, 3/2 +}^{-1} \otimes \ket{\nu 200\, 1/2 -}^1$  &$-$181.50   & 12.83 &   0.34    &  $2_Z$\\
10  &  $\ket{\nu 220\, 1/2 +}^{-1} \otimes \ket{\nu 211\, 1/2 -}^1$  &$-$181.99   & 12.34 &   0.35  &  $1_Z$\\
11  &   $\ket{\nu 220\, 1/2-}^{-1} \otimes \ket{\nu 211\, 1/2-}^1$  &$-$180.78   & 13.55 &   0.33  &   $0_Z$\\
12  &   $\ket{\nu 211\, 3/2-}^{-1} \otimes \ket{\nu 202\, 3/2+}^1$  &$-$178.83   & 15.50 &   0.34  &   $3_Z$\\
13  &   $\ket{\nu 211\, 3/2+}^{-1} \otimes \ket{\nu 202\, 3/2+}^1$  &$-$177.16   & 17.17 &   0.33  &   $0_Z$\\
14  &  $\ket{\nu 220\, 1/2 -}^{-1} \otimes \ket{\nu 200\, 1/2 -}^1$  &$-$177.04   & 17.29 &   0.27   &  $0_Z$\\
15  &   $\ket{\nu 220\, 1/2+}^{-1} \otimes \ket{\nu 200\, 1/2-}^1$  &$-$176.94   & 17.39 &   0.25   &   $1_Z$\\
16  &   $\ket{\nu 220\, 1/2-}^{-1} \otimes \ket{\nu 202\, 3/2+}^1$  &$-$174.00   & 20.33 &   0.25   &   $2_Z$\\
17  &   $\ket{\nu 211\, 3/2+}^{-1} \otimes \ket{\nu 202\, 3/2+}^1$  &$-$173.47   & 20.86  &  0.24  &   $1_Z$\\
18  &  $\ket{\pi 211\, 3/2 -}^{-1} \otimes \ket{\pi 202\, 5/2 -}^1$  &$-$188.00   & 6.33 &   0.34  &  $1_Z$\\
19  &   $\ket{\nu 211\, 3/2-}^{-1} \otimes \ket{\nu 202\, 5/2-}^1$  &$-$184.29   & 10.04  & 0.10   &   $1_Z$\\
 \mbox{}  &   $\ket{\pi 211\, 3/2-}^{-1} \otimes \ket{\pi 202\, 5/2-}^1$  & \mbox{}   &   \mbox{}       &     \mbox{}  & \mbox{}\\
20  &   $\ket{\nu 211\, 3/2}^{-2} \otimes \ket{\nu 202\, 5/2}^2$  &$-$183.13   &      11.20  &  0.26   &   $0_Z$\\

\end{tabular}
\end{ruledtabular}
\end{table}
\nopagebreak[4]

\subsection{Configuration space}

The configuration spaces for $^{24}$Mg is built by following the general rules sketched in Sect.~\ref{sec:ncci}. 
We include the ground state and all possible 1p-1h excitations among active $N=2$ Nilsson 
levels shown in Fig.~\ref{fig:studnia}. In self-conjugated nuclei the isospin projection allows to reduce the space by considering
$p-h$ excitations of a single, say neutron, charge. 
Indeed, the proton s.p. levels are almost identically spaced and just pushed higher in energy due to the Coulomb interaction.
Simple counting shows that there is 16 different $\nu p-\nu h$ excitations. In addition, we include in the configuration spaces  two lowest 2p-2h configurations. These configurations are added in order to test stability 
of the GTSD with respect to higher order excitations. All HF states included in the configuration space of $^{24}$Mg are listed in Table~\ref{tab:confsd}. They are prolate deformed axially 
symmetric configurations.

\begin{figure}[htb]
\centering
\includegraphics[width=0.7\columnwidth]{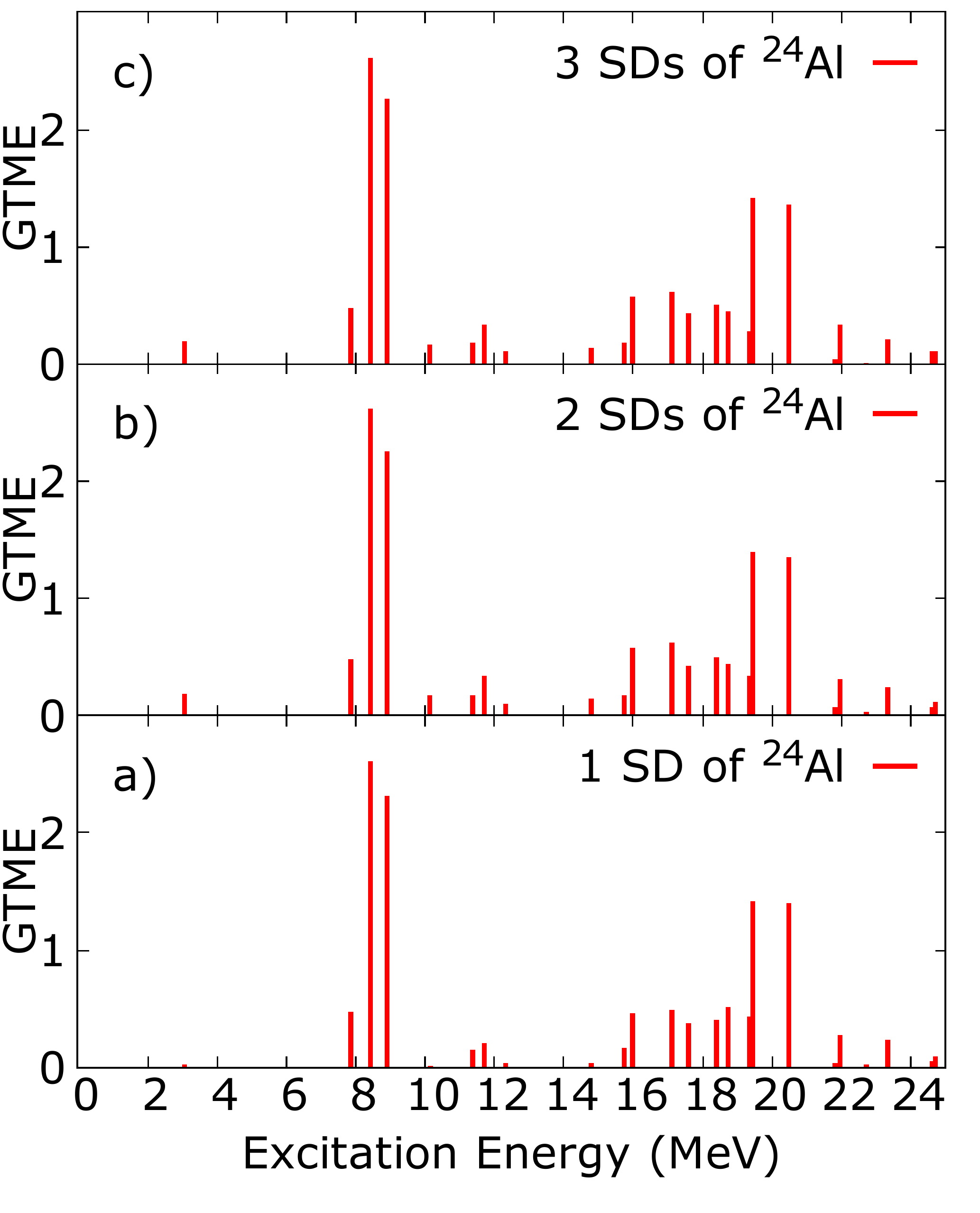}
\caption{(Colour online) GTME stability analysis against configuration mixing in the parent nucleus. The figure shows GTMEs  for
$|^{24}\textrm{Al}; 4_\textrm{g.s.}^+\rangle \to  | ^{24}\textrm{Mg}; 4^+ \rangle$ decay.
The results were obtained by using 17 configurations in the daughter nucleus. The number of Slater determinants (SDs) used to correlate the parent nucleus
changes from one (bottom) to three (top).}
\label{fig:parent}
\end{figure}

A systematic study of the GT matrix elements (GTMEs) in $T=1/2$ mirror nuclei~\cite{(Kon16)} allowed us to conclude 
that these g.s. to g.s. $I^\pi \rightarrow I^\pi $ matrix elements are fairly insensitive on the configuration mixing. Similar 
property appears to hold here as demonstrated in Fig.~\ref{fig:parent}. The figure shows calculated 
$4^+ \rightarrow 4^+$ GTMEs between the g.s. of $^{24}$Al and the $4^+$ states in $^{24}$Mg
calculated by using the DFT-NCCI model with 17 configurations involving the g.s. and all 1p-1h excitations. Each of the panel differs by the treatment of the g.s. DFT-NCCI wave function of the parent nucleus. 
We start with the wave function projected from  the so called aligned SR g.s. configuration (a) and enrich it by admixing first the anti-aligned g.s. (b) and, eventually, the lowest  1p-1h excitation (c). 
As clearly visible, the calculated GTMEs are almost insensitive to the wave function in the parent nucleus. Hence, in all calculations shown below, 
the correlated g.s. DFT-NCCI wave function in $^{24}$Al includes three SR Slater determinants: 
the aligned g.s., the anti-aligned g.s., and the lowest 1p-1h excitation.   

\begin{figure}[htb]
\centering
\includegraphics[width=0.8\columnwidth]{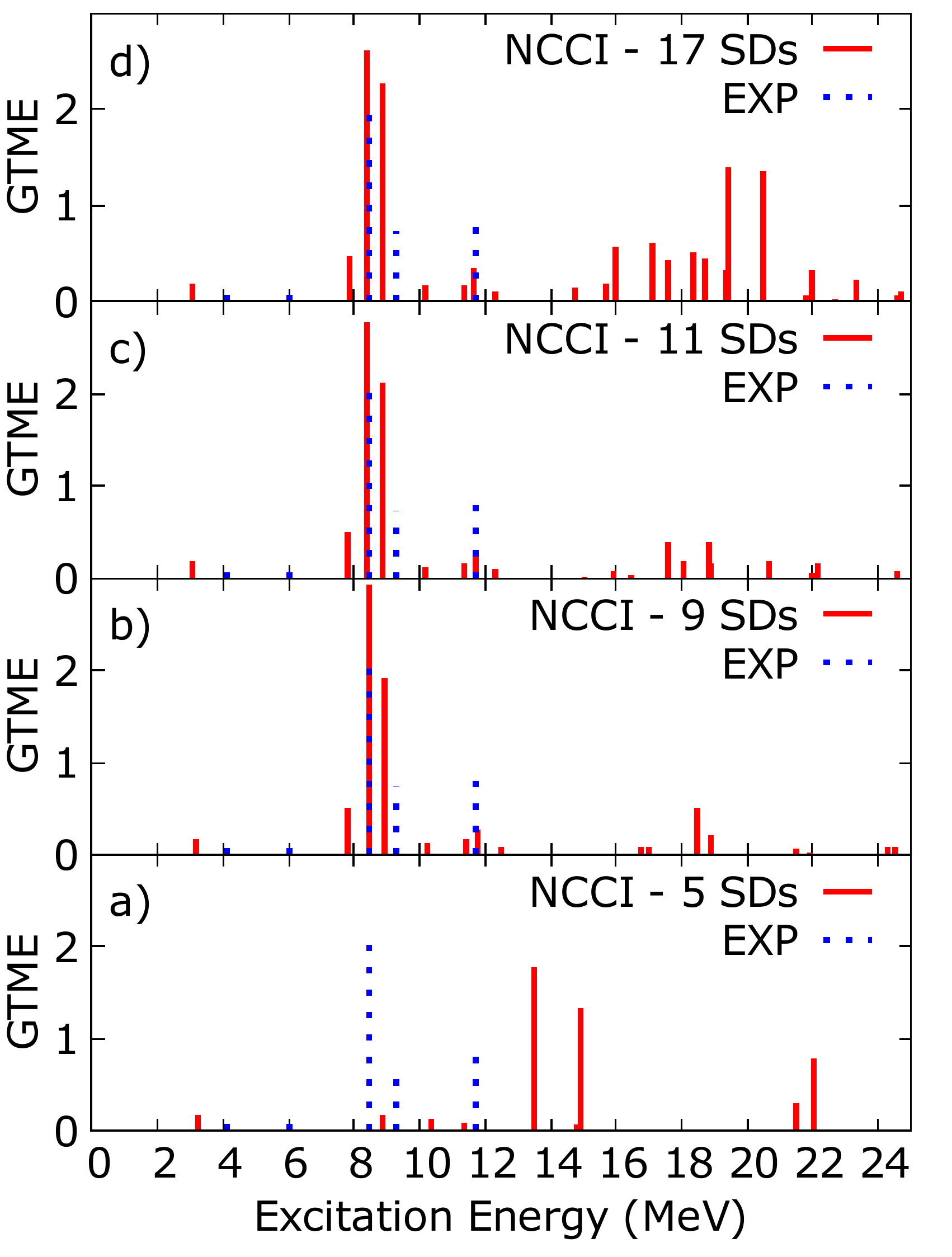}
\caption{(Colour online) GTMEs  between the $4^+$ g.s. of $^{24}$Al and the $4^+$ states in $^{24}$Mg for various dimensions of the used configuration space in $^{24}$Mg. See text for details.}
\label{fig:24Mg}
\end{figure}

\subsection{Gamow-Teller strength distribution}

In order to pin down a specific role played by different Nilsson levels we studied diagonal $I^\pi \to I^\pi$ GTMEs
as a function of the configuration space size in the daughter nucleus. The results are illustrated in Fig.~\ref{fig:24Mg}. The bottom panel (a) shows the GTME distribution  calculated using HF configurations involving the g.s. and all 1p-1h excitations among the Nilsson levels originating from the spherical $d_{5/2}$
sub-shell. Panel (b) contains additionally all 1p-1h configurations involving the $\ket{211\, 1/2}$ Nilsson state originating from
the spherical $d_{3/2}$ sub-shell. This level plays critical role in shaping up the GTSD in $^{24}$Mg around the excitation energy 
of $\sim$8\,MeV. This example shows how sensitive the GTSD is with respect to
the position of s.p. levels.
Panel (c) shows the GTMEs calculated using all HF configurations having excitation energies, $\Delta E_{\text{HF}}$,
below the experimental $Q_\beta$ value i.e. $\Delta E_{\text{HF}} \leq 14.5$\,MeV.  Eventually, (d) shows the distribution calculated using all available $p-h$ configurations.
The configurations included in the panels (c) and (d) influence predominantly the high-energy part of GTME distribution,  
above the experimental $Q_\beta$ window.

\begin{figure}[htb]
\centering
\includegraphics[width=1.0\columnwidth]{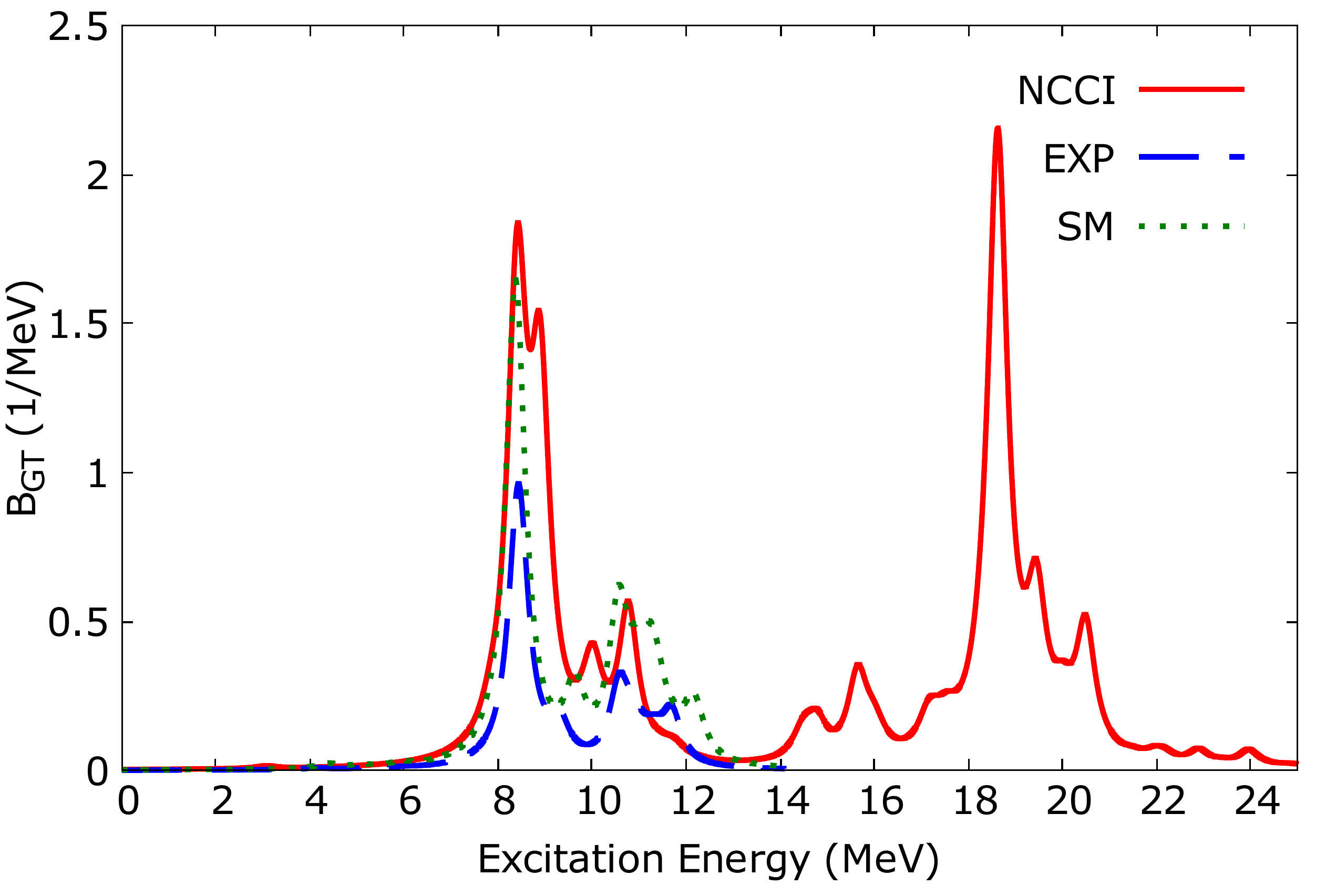}
\caption{(Colour online) Complete GTSD in $^{24}$Mg following beta decay of  the $4^+$ g.s. in $^{24}$Al. The GTSD includes matrix 
elements to $3^+,4^+$, and $5^+$ states in $^{24}$Mg.  The DFT-NCCI result, is compared to experimental 
data, and the shell-model calculations of Ref.~\cite{(Bro85a)}. Curves are smoothed with the Lorentzian function 
with half-width of $\Gamma=0.5$\,MeV.}
\label{fig:24MgSD}
\end{figure}

The result from the DFT-NCCI calculation, including transitions from the $4^+$ g.s. of $^{24}$Al to 
all $3^+,4^+$, and $5^+$ states in $^{24}$Mg, is shown in Fig.~\ref{fig:24MgSD}. The calculated GTSD is compared to 
USDb shell-model calculation and experiment \cite{(Bro85a)}. Both the shell model and DFT-NCCI results are in a perfect 
agreement with experiment concerning position of a centroid describing $4^+\to 4^+$ transition but the theoretical peaks are roughly two 
times higher compared to experiment. Moreover, in the DFT-NCCI calculations the first resonant peak, called hereafter the first GT resonance (GTR1), 
splits into two close-lying peaks.  The second GT resonance (GTR2)  seen in the DFT-NCCI calculation at high excitations energies 
is well above the experimental $Q_\beta$-energy. 

\begin{figure}[htb]
\centering
\includegraphics[width=1.0\columnwidth]{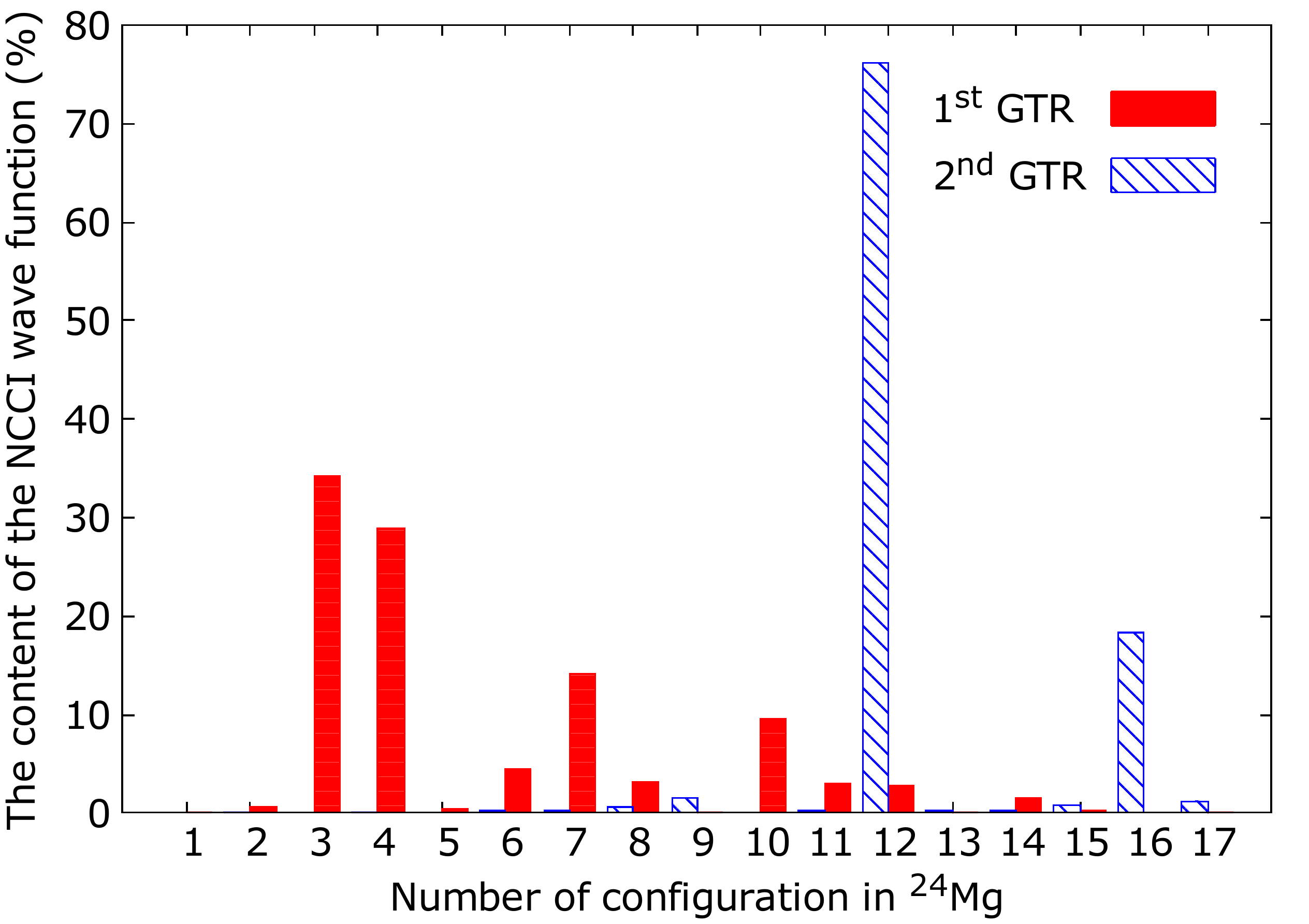}
\caption{(Colour online) The content of the wave function of first $4^+$ Gamow$-$Teller resonance (red) and of the second GTR peak in terms of HF configurations listed in Table~\ref{tab:confsd}}
\label{fig:GTGRmg}
\end{figure}

In order to reveal the nature of resonant transitions, their wave functions has been decomposed 
in terms of HF configurations, shown in Fig.~\ref{fig:GTGRmg}.  The $4^+$ g.s. of  $^{24}$Al is dominated by the aligned HF configuration 
having an unpaired proton on $\ket{202\, 5/2}$ s.p. level.  This level has a large GT s.p. matrix elements with the $\ket{202\, 5/2}$ level and its 
spin-orbit partner, $\ket{202\, 3/2}$, in $^{24}$Mg.  Hence, the GTR1 is due to transition to the aligned $p-h$ excitation involving neutron particle in 
$\ket{202\, 5/2}$. Its structure, however,  is strongly affected  by the aligned $p-h$ excitation involving neutron particle in 
$\ket{211\, 1/2}$ orbit due to proximity of the $\ket{202\, 5/2}$ and $\ket{211\, 1/2}$ levels in
the potential well, see Fig.~\ref{fig:studnia}. The near-degeneracy causes mixing between the states projected from 
these HF configurations  since the $K$ quantum number is not conserved.

\begin{figure}[htb]
\centering
\includegraphics[width=1.0\columnwidth]{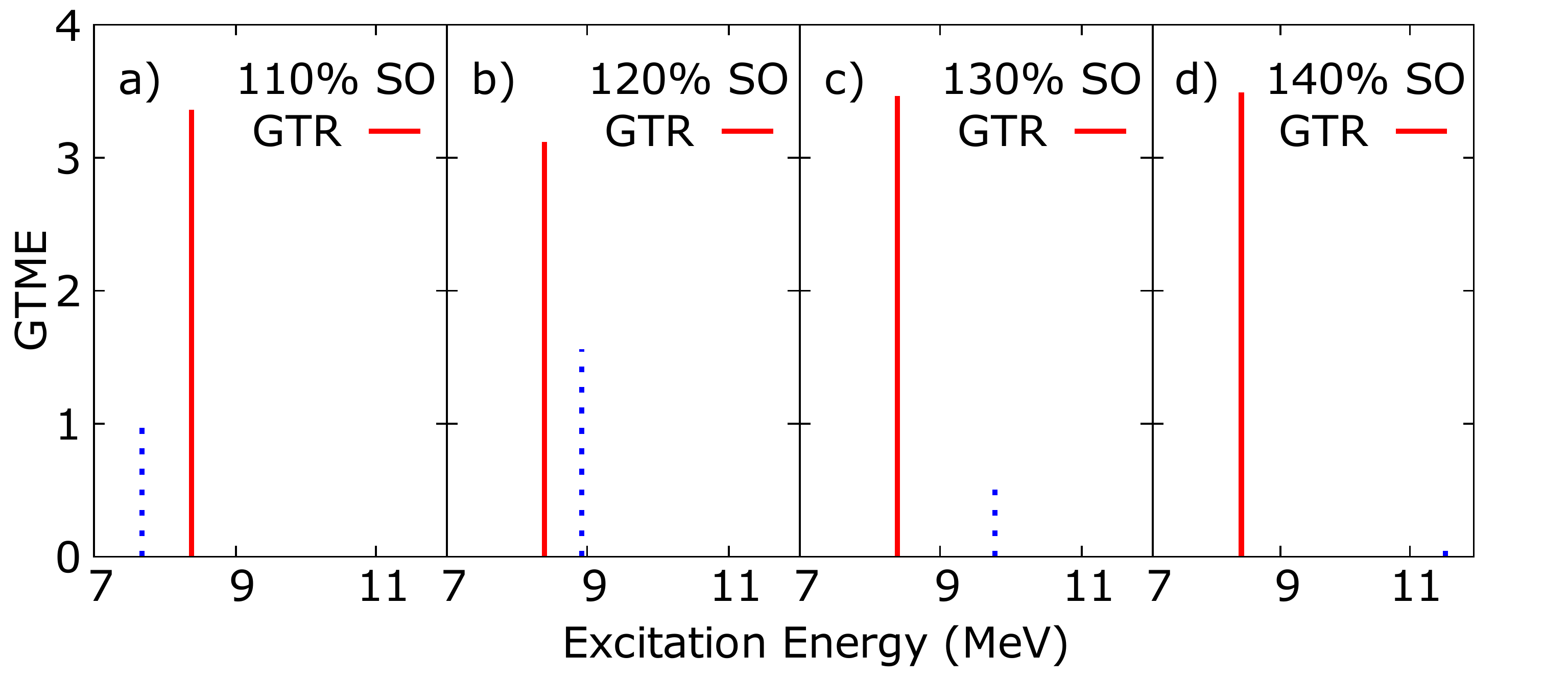}
\caption{(Colour online) GTMEs connecting the $4^+$ g.s. in $^{24}$Al with the $\ket{202\, 5/2}$ resonant peak (solid line) and 
the  $\ket{211\, 1/2}$ secondary peak (dotted line). The calculations were done  using the SV$_{\rm T}$ parameterization with spin-orbit strength
increased by $10\%$ (a), $20\%$ (b), $30\%$ (c), and $40\%$ (d) with respect to the original value. See text for further
details.}
\label{fig:so24mg}
\end{figure}

By slightly increasing the spacing between the $\ket{202\, 5/2}$ and $\ket{211\, 1/2}$
s.p. levels, the mixing becomes reduced, which increases purity of the GTR1 wave function and further improves agreement 
with the data.  The spacing can be increased, for example, by slight increase of the spin-orbit strength. The result of such a test study is 
shown in Fig.~\ref{fig:so24mg}.  The figure shows a series of DFT-NCCI calculations using the SV$_\textrm{T}$ Skyrme force with the 
spin-orbit strength increased by $10\%$, $20\%$, $30\%$, and $40\%$ with respect to the original value.  The configuration space 
in $^{24}$Mg used in this test study was constrained to the SR g.s. and two aligned $p-h$ excitations to $\ket{202\, 5/2}$ and 
$\ket{211\, 1/2}$ levels.  The calculation shows that the centroid of the main peak and its height weakly depends 
on the spin-orbit strength at variance to the secondary peak associated with the $\ket{211\, 1/2}$ Nilsson level. 
Indeed, with increasing the spin-orbit strength the secondary peak moves toward higher energies and its magnitude decreases. 
The study suggest that the optimal spin-orbit strength should be around 25\% larger than the original value. 
At contrast to the GTR1, the structure of GTR2 is predicted to be very pure with 80\% of its wave function content coming from the aligned 
1p-1h excitation involving neutron particle in the $\ket{202\, 3/2}$ Nilsson level.

\section{Superallowed Gamow-Teller beta decay of $^{100}$Sn}\label{sec:A100}

In this section we have computed the superallowed Gamow-Teller beta decay of the heaviest $N=Z$ nucleus $^{100}$Sn 
and the low-spin structure, $I\leq 8$, in the daughter nucleus $^{100}$In. The transition, which proceeds from the 
$0^+$ g.s. of $^{100}$Sn to  the first $1_1^+$ state in $^{100}$In, is the fastest GT decay observed so far, see 
Ref.~\cite{(Hin12)}. The GTME is well reproduced by the dedicated Large-Scale-Shell-Model calculations 
under the assumption that the  axial coupling constant is quenched by 40\%~\cite{(Hin12)}.

The aim is to test the universality of DFT-NCCI approach which, at least {\it in principle\/}, can be 
applied to calculate both the nuclear spectra and transition rates  in any atomic nucleus, 
irrespectively of its mass and particle-number parity.  
Hence, in the calculation we have used exactly the same formalism as in the preceding sections. The HF configurations 
were calculated by using the SV$_{\textrm{SO}}$ variant of the Skyrme SV$_{\textrm{T}}$ force within the space consisting of 12 spherical 
harmonic oscillator shells. 

The SV$_{\textrm{SO}}$ has a 20\% stronger spin-orbit interaction strength compared to SV$_{\rm T}$~\cite{(Kon16)}.
As discussed in Ref.~\cite{(Kon16)}, the use of  SV$_{\textrm{SO}}$ variant considerably improves calculated masses 
in $N\sim Z$ nuclei as compared to the DFT-NCCI calculations based on the SV$_{\textrm{T}}$ force. 
  
In case of the doubly-magic $^{100}$Sn we considered only a single mean field configuration 
representing its g.s. The calculated binding energy 827.7\,MeV of  $^{100}$Sn is in a fair agreement with the experimental 
value 825.3$\pm$0.3\,MeV overestimating it by 0.3\%.

The structure of $^{100}$In was computed by using nine axially-deformed mean field configurations. Counting with respect to the $^{100}$Sn core, 
eight of them correspond to p-h configurations with the neutron particle occupying different s.p. states 
originating from the $d_{5/2}$ and $g_{7/2}$ spherical sub-shells and  the proton hole being in the s.p. orbital 
originating from $g_{9/2}$ spherical sub-shell.  In spherical language these are $\nu d_{5/2}\otimes \pi g_{9/2}^{-1}$
and $\nu f_{5/2}\otimes \pi g_{9/2}^{-1}$ configurations. The ninth configuration, involving the lowest 
$\pi$p-$\pi$h excitation through the $Z$=50 shell gap, was added to test stability against the cross-shell excitations. 
The calculation shows that it does not affect neither the low-lying spectrum nor the GTME.

\begin{figure}[htb]
\centering
\includegraphics[width=1.0\columnwidth]{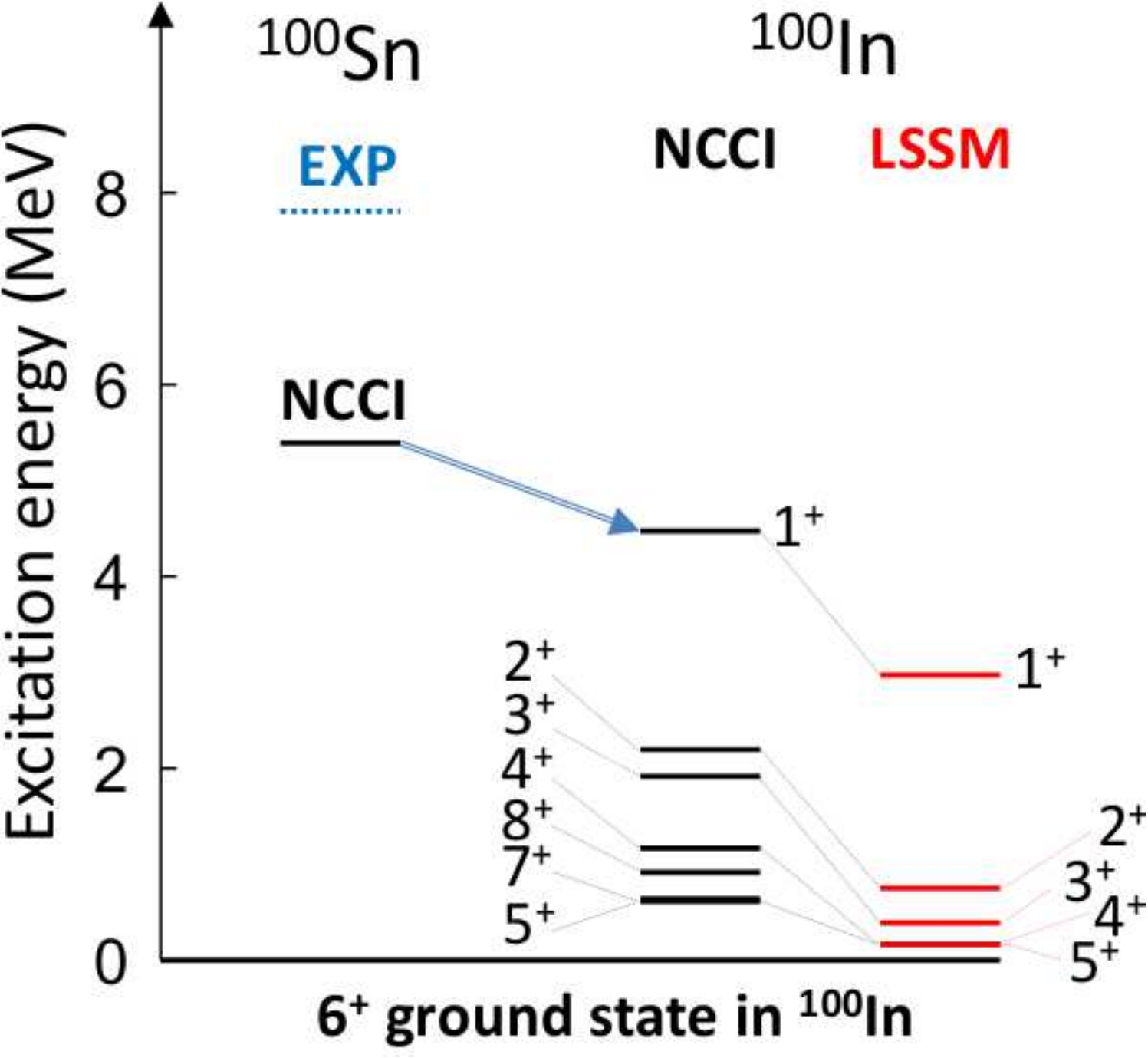}
\caption{(Colour online)  
Low-lying states in $^{100}$In calculated by using the DFT-NCCI  (middle) and LSSM (right).
Left part shows the experimental (dotted line) and DFT-NCCI (solid line) binding energies in  $^{100}$Sn 
relative to the ground state energy in $^{100}$In. See text for details.}
\label{fig:100sn}
\end{figure}

The calculated spectrum, which includes the first  $1^+\leq I^\pi \leq 8^+$ states in $^{100}$In, is depicted in Fig.~\ref{fig:100sn} and 
compared  to the LSSM results for the first $1^+\leq I^\pi \leq 6^+$, taken from Ref.~\cite{(Hin12)}. We refrain from showing the 
experimental spectrum since neither the spins nor the excitation energies are firmly assigned~\cite{(Hin12)}.  Theoretical 
spectra were normalized to the g.s. energy which is predicted to have $I=6^+$ by both the models. 
The predicted DFT-NCCI binding energy for this state is in perfect agreement with the experimental binding 
energy of  $^{100}$In underestimating it only by 9\,keV. Concerning excited states, the DFT-NCCI model predicts the following   
values: 0.618\,MeV ($5_1^+$),  0.637\,MeV ($7_1^+$), 0.927\,MeV ($8_1^+$), 1.176\,MeV ($4_1^+$), 
1.912\,MeV ($3_1^+$), 2.194\,MeV  ($2_1^+$), and 4.475\,MeV  ($1_1^+$). The excitation energy of $8^+$ state may be  
somewhat uncertain due to too small number of knots used in the integration over the   angles in the angular-momentum projection procedure. The level ordering agrees relatively well with the LSSM
calculations but the excitation energies are systematically larger.  Note, that the DFT-NCCI model predicts the low-lying doublet 
composed of  $5_1^+$ and $7_1^+$ states at variance to the LSSM which predicts near-degeneracy of the first  $5_1^+$ and
$4_1^+$ states. 

The calculated B$_\textrm{GT}^\textrm{(NCCI)} \approx 10.2$ after using the effective axial-vector 
strength, $g_\textrm{A}^{\textrm{(eff)}} = q  g_\textrm{A}$,   quenched by 40\% \cite{(Hin12)} with respect to the free-neutron 
value of $g_\text{A} = -1.2701$. The quenching factor $q$=0.6 is typical for $A\approx 100$ mass region~\cite{(Pir15)}. The quenched 
B$_\textrm{GT}^\textrm{(NCCI)}$ agrees well with the experimental value   B$_\textrm{GT}^\textrm{(EXP)} = 9.1^{+2.6}_{-3.0}$.

\section{Conclusions}\label{sec:conc}
 
In the present work
we have presented pioneering calculations of the Gamow-Teller transitions by using the no-core-configuration-interaction
approach based on multi-reference density functional theory treating properly the isospin 
and rotational symmetries.  The DFT-NCCI formalism was applied to compute the GTSD in the $p$-shell $^8$Li and $^8$Be nuclei. Although the model lacks the coupling to continuum essential to describe broaden resonances and in turn beta-decay properties, we have shown that it can provide an input to theories exploring open-channel physics such as R-matrix. Shell-model calculation applied to such approach supported experimental-data analysis. It may be of particular interest to follow the path with entirely different DFT-rooted theory.    
Moreover, we have demonstrated that the model is capable to capture the GTSD satisfactorily well using relatively small {\it configuration space\/} in the $sd$-shell $^{24}$Mg as well. It was also shown that the model allows 
for interpretation of the GTSD peaks in terms of specific Nilsson orbits of deformed mean field,
i.e. in a way that is complementary to the traditional nuclear shell model calculations.

The DFT-NCCI model can be, at least {\it in principle\/}, applied to calculate both the nuclear spectra and transition 
rates in atomic nuclei irrespectively of their mass and particle-number parity.  In order to demonstrate its flexibility, 
the model was also applied to compute the superallowed GT beta decay in $^{100}$Sn and the low-spin spectrum 
in $^{100}$In. It is shown that, after applying the standard quenching factor of $q\approx 0.6$, the calculated matrix 
element agrees well with the experimental value. The low-spin spectrum agrees quite well with 
the large-space-shell-model calculation of Hinke~\cite{(Hin12)}. Eventually, let us stress that all the results 
presented above were obtained without any readjustment of the model parameters to experimental data.

In conclusion, we have demonstrated that DFT-NCCI formalism can be successfully used to study nuclear beta decay in diverse set of nuclei, 
thus offering a complementary method to {\it ab initio\/} and shell model
approaches. This study paves a way for more systematic studies of nuclear beta decay rates, 
for exploring forbidden beta-decays, and for tackling the double-beta-decay process within the DFT-NCCI framework. 
With forbidden beta-decays, the spectrum-shape method may offer valuable hints for the $g_{\rm A}$ quenching puzzle~\cite{(Haa16)}.
Although the correspondence to experimental results was generally found to be rather good, the underlying effective interaction
used to construct the EDF has its limitations. A work towards developing novel EDFs applicable for beyond-mean-field
calculations is under way.

\begin{acknowledgments}

This work was supported by the Polish National Science Center (NCN) under Contract Nos. 2014/15/N/ST2/03454 and 2017/24/T/ST2/00160, by the Academy of Finland under the Centre of Excellence
Programme 2012-2017 (Nuclear and Accelerator Based Physics Programme at JYFL) and by the FIDIPRO program.
We acknowledge the CIS-IT National Centre for Nuclear Research (NCBJ), Poland and CSC-IT Center for Science Ltd, Finland for allocation of computational resources.   

\end{acknowledgments}

\bibliography{NCCI,jacwit33}

\end{document}